\documentclass[11pt,a4paper]{article}
\usepackage{float}
\usepackage[T1]{fontenc}
\usepackage[utf8]{inputenc}
\usepackage{lmodern,amsmath,amssymb,amsthm,mathtools,bm}
\usepackage[margin=28mm]{geometry}
\usepackage{authblk,booktabs,array,graphicx,tikz}
\usepackage{enumitem,placeins}
\setlist{itemsep=4pt,topsep=5pt}
\allowdisplaybreaks[2]
\usepackage[numbers,sort&compress]{natbib}
\definecolor{linkgreen}{RGB}{0,110,60}
\usepackage[colorlinks=true,linkcolor=linkgreen,citecolor=linkgreen,urlcolor=linkgreen]{hyperref}
\usepackage{microtype}
\hypersetup{pdftitle={Spectral Disks for Quantum Markov Generators: Escape Rates and Optimal Representations},pdfauthor={D. Chruscinski, S. Denisov, W. Tarnowski, K. Zyczkowski}}
\newtheorem{theorem}{Theorem}[section]
\newtheorem{proposition}[theorem]{Proposition}
\newtheorem{corollary}[theorem]{Corollary}
\newtheorem{lemma}[theorem]{Lemma}
\theoremstyle{definition}\newtheorem{example}[theorem]{Example}
\newtheorem{remark}[theorem]{Remark}
\newcommand{\tr}{\operatorname{tr}}
\newcommand{\Tr}{\operatorname{Tr}}
\newcommand{\id}{\operatorname{id}}
\newcommand{\spec}{\operatorname{spec}}
\newcommand{\CP}{\mathrm{CP}}

\newcommand{\cL}{\mathcal L}
\newcommand{\D}{\mathcal D}

\newcommand{\ket}[1]{|#1\rangle}
\newcommand{\bra}[1]{\langle#1|}

\title{Spectral Disks for Quantum Markov Generators}
\author[1]{Dariusz Chru\'sci\'nski}
\author[2]{Sergey Denisov}
\author[3]{Wojciech Tarnowski}
\author[3,4]{Karol \.Zyczkowski}
\affil[1]{Institute of Physics, Nicolaus Copernicus University, 87-100 Toru\'n, Poland}
\affil[2]{Department of Computer Science, Oslo Metropolitan University, N-0130 Oslo, Norway}
\affil[3]{Institute of Theoretical Physics, Jagiellonian University, 30-348 Cracow, Poland}
\affil[4]{Center for Theoretical Physics, Polish Academy of Sciences, 02-668 Warsaw, Poland}
\begin{document}
\maketitle
\begin{abstract}
Spectral bounds for Markov generators require a choice of time scale. In quantum dynamics, a rate read off
from a jump representation, the escape rate from a pure state and the smallest spectral disk need not agree.
We prove that the spectrum of every GKLS generator written without a Hamiltonian lies in a disk tangent to the
imaginary axis, with radius given by the norm of the operator $\Phi^\dagger(I)$ built from the jump map $\Phi$.
This settles the first two scaling conjectures of \emph{Corralling spectra of generators of time-continuous
Markov evolution (both classical and quantum)}, J. Phys. A: Math. Theor. \textbf{59}, 175202 (2026).
The third, escape-rate conjecture fails for a qutrit with a single normal jump. Nevertheless, twice the escape
rate bounds every decay rate of any positive trace-preserving semigroup For GKLS generators with vanishing canonical Hamiltonian, four times the escape rate always supplies a spectral disk, and the factor four is optimal uniformly in dimension. We compare these bounds with classical Cassini ovals and thermodynamic ellipses,
explain the obstacles to their quantum extensions, and formulate optimization over completely positive
jump-map representations as a semidefinite program.
\end{abstract}


\section{Introduction}
How rapidly can a Markovian quantum system relax, and how fast can its decaying modes oscillate?
Both questions concern the spectrum of the generator. An eigenvalue $-\Gamma+i\omega$ describes a mode with
decay rate $\Gamma$ and oscillation frequency $\omega$. Positivity and trace preservation place this spectrum
in the closed left half-plane, but this qualitative constraint does not compare systems with different
interaction strengths. 

For a finite classical Markov chain there is a natural answer: the largest exit rate from an individual state.
Gershgorin's theorem then confines the spectrum to a disk tangent to the imaginary axis at the origin, 
that is, if $K$ is the matrix of a classical Markov generator then $\spec(K) \subseteq D(-\nu_c,\nu_c)$, where $\nu_c = \max_i |K_{ii}|$ and 

\begin{equation}
D(z_0,r)=\{z\in\mathbb C:|z-z_0|\le r\},\qquad r\ge 0,
\end{equation}
defines a closed disk on the complex plane. 

%\cite{corralling}. 
%The same rate permits uniformization, expressing the continuous-time generator as a multiple of a stochastic
%matrix minus the identity. These two facts make the classical normalization both spectrally useful and
%probabilistically transparent. Quantum dynamics separates them. Already for amplitude damping the Euler step
%$\id+\mathcal L/\nu$ is not positive for any finite $\nu>0$, although its spectrum satisfies a simple disk %bound.
%The absence of classical uniformization therefore does not settle the quantum spectral problem.

In a recent paper 
%\emph{Corralling spectra of generators of time-continuous Markov evolution (both classical %and quantum)}
\cite{corralling}, we analyzed this problem for purely dissipative quantum Markov generators represented by the celebrated Gorini-Kossakowski-Lindblad-Sudarshan (GKLS) form \cite{gks,lindblad}

\begin{equation}\label{eq:gkls}
\cL(X)=-i[H,X]+\Phi(X)-\frac12\{G,X\}
\end{equation}
where the \emph{jump map} $\Phi$ and the operator $G$ are
\begin{equation}
\Phi(X)=\sum_\alpha  V_\alpha XV_\alpha^\dagger,\quad
G=\Phi^\dagger(I)=\sum_\alpha V_\alpha^\dagger V_\alpha .    
\end{equation}
%For a single jump $V$ we write $\D_V(X)=VXV^\dagger-\frac12\{V^\dagger V,X\}$. The disk theorem below concerns a specified representation with $H=0$. 
In a {\it canonical representation} the jumps $V_\alpha$ are traceless mutually orthogonal 
$\tr V_\alpha^\dagger V_\beta=\gamma_\alpha \delta_{\alpha\beta}$ and we put a canonical Hamiltonian $H=0$ (purely dissipative regime). In \cite{corralling} we introduced three {\em quantum scales}

\begin{align}
\nu_{q,1}&=\sum_\alpha\gamma_\alpha = \frac 1n\, {\rm Tr}(- \cL) ,\label{eq:nu1}\\
\nu_{q,2}&=\|G\|_\infty,\label{eq:nu2}\\
\nu_{q,3}&=\max_{\|\psi\|=1}\left|\bra\psi\cL(\ket\psi\bra\psi)\ket\psi\right| . \label{eq:nu3}
\end{align}
They measure, respectively, the total canonical GKLS
rate, the largest total jump intensity, and the largest instantaneous escape rate from a pure state.
These quantities describe different aspects of the dynamics. Conjecture $i$ of \cite{corralling} ($i=1,2,3$) asserts that the spectrum of $\cL$ is confined to the disk determined by $\nu_{q,i}$, and the conjectures were accompanied by the ordering 

\begin{equation} \label{123}
    \nu_{q,3} \leq \nu_{q,2} \leq \nu_{q,1} ,
\end{equation}
of the three scales. Explicitly, the proposed inclusions read

\begin{equation} \label{L-123}
    \spec(\cL) \subseteq D(-\nu_{q,i},\nu_{q,i}) , \quad i\in \{1,2,3\} .    
\end{equation}
Clearly, (\ref{123}) implies the following hierarchy of nested disks

\begin{equation}
     D(-\nu_{q,3},\nu_{q,3})    \subseteq D(-\nu_{q,2},\nu_{q,2}) \subseteq D(-\nu_{q,1},\nu_{q,1}) .
\end{equation}
These conjectures were supported by several solvable models and numerical analysis \cite{corralling}. 

In this paper we prove the ordering (\ref{123}) and Conjectures 1 and 2, i.e.\ we show that
(\ref{L-123}) holds for $i\in\{1,2\}$. The proof rests on a disk theorem for GKLS generators,
which extends to generators built from a merely positive jump map. In general, however,
$\spec(\cL)\not\subseteq D(-\nu_{q,3},\nu_{q,3})$: Conjecture 3 is violated, and we present a
qutrit counterexample with a single normal jump, communicated to us by W.~Zhang \cite{Zhang}.
Conjecture 3 nevertheless holds for every qubit generator with vanishing canonical Hamiltonian.
For arbitrary positive trace-preserving semigroups we prove the sharp decay-rate bound
$\Gamma_{\max}\le2\nu_{q,3}$, without assuming unitality or complete positivity
(Theorem~\ref{thm:universal-decay-rate}). Consequently, every real eigenvalue belongs to the
escape-rate disk, and the full disk inclusion holds whenever the spectrum is real.
This includes Hilbert--Schmidt self-adjoint generators and the detailed-balance setting in
which the entire Heisenberg generator is self-adjoint in a weighted inner product associated
with a faithful state. The qutrit counterexample shows that controlling decay rates alone does
not control the oscillation frequencies sufficiently to establish Conjecture~3.
For GKLS generators with vanishing canonical Hamiltonian, however, a larger universal disk is
available: $R(\cL)\le\nu_{q,2}\le4\nu_{q,3}$, with the factor four optimal uniformly over all
finite dimensions (Theorem~\ref{thm:sharp4}).

Classically, the exit-rate disk is only the coarsest of several known spectral enclosures.
Brauer's Cassini ovals \cite{Brauer1946, Brauer1947, Varga2004, Krishna2026} sharpen it using the two largest exit rates. Two thermodynamic ellipses constrain the
oscillatory part of the spectrum by irreversibility: the edge-force ellipse of \cite{Xu2025} bounds the spectrum,
and even a weighted numerical range~\cite{HornJohnson2013}, through the largest thermodynamic force on a transition, while the
Uhl--Seifert ellipse \cite{UhlSeifert2019} uses cycle affinities; a proof of the latter has recently been
proposed \cite{Kolchinsky2026}. At (classical) detailed balance both ellipses collapse to a real segment. We collect these
refinements, give short proofs of the steps we use, and show why they do not pass to quantum generators through
a fixed classical compression. Coherence modes carry eigenvalues that population dynamics does not see; already
the Cassini oval fails for the quantum embedding of a classical generator, and the qutrit counterexample excludes
every ellipse tangent to the imaginary axis at the origin whose horizontal semiaxis is the escape rate. The real-spectrum results above describe the reversible
endpoint of this picture.

A second distinction concerns representation freedom. The spectrum and the escape rate depend only on
$\mathcal L$, whereas the operator $G$ depends on how the generator is written. Adding an identity jump changes
the total jump intensity without changing the evolution. Canonical traceless jumps remove this ambiguity in
a prescribed way, but need not minimize the norm of $G$. We therefore optimize over all completely
positive jump-map representations of the fixed generator, which leads to a semidefinite program.
Explicit examples show that its optimum can exceed the exact spectral scale.

The resulting picture distinguishes three questions: which disk contains the spectrum, how rapidly a pure
state initially loses survival probability, and how small a completely positive jump-map representation can be made.
They agree for some important families and differ for others.

Section~\ref{sec:disk} proves the disk theorem, its classical counterpart, its
Hamiltonian enlargement, and Conjectures 1 and 2. Section~\ref{sec:false} examines Conjecture 3: the qutrit
counterexample, the sharp dimension-independent factor-four disk, and the universal decay-rate bound
with its real-spectrum and detailed-balance consequences. Section~\ref{sec:optimization} develops
representation optimization, including a semidefinite program and exactly solvable examples.
Section~\ref{sec:cassini-thermo} compares the disk with Brauer's Cassini ovals and the thermodynamic ellipses
of classical generators, and
Section~\ref{sec:conclusions} draws the main consequences and formulates open problems.

%Appendix~\ref{app:lemma} extends the disk theorem to positive jump maps, and
%Appendix~\ref{sec:geometry} analyzes a one-parameter family of qutrit counterexamples. %Appendix~\ref{app:inside} gives an inner disk that contains no eigenvalue of a single-jump generator, so that %the spectrum lies in a ring.

\section{A disk theorem for GKLS generators}\label{sec:disk}

The classical Gershgorin argument selects a component of an eigenvector having maximal modulus. We first record the exact disk scale of a given spectrum.

\begin{proposition}[Exact spectral scale]\label{prop:R}
Suppose every nonzero eigenvalue of $\cL$ has strictly negative real part. Then the smallest $\nu\ge0$ for which $\spec(\cL)\subseteq D(-\nu,\nu)$ is
\begin{equation}\label{eq:R}
R(\cL)=\max_{\lambda\in\spec(\cL)\setminus\{0\}}
\frac{|\lambda|^2}{-2\Re\lambda},
\end{equation}
where the empty maximum is zero.
\end{proposition}
\begin{proof}
For $\nu>0$, $|\lambda+\nu|^2\le\nu^2$ is equivalent to $|\lambda|^2+2\nu\Re\lambda\le0$, i.e.\ to $\nu\ge|\lambda|^2/(-2\Re\lambda)$ when $\Re\lambda<0$. This gives \eqref{eq:R}. For $\nu=0$ the disk is $\{0\}$, which contains the spectrum exactly when it is $\{0\}$, i.e.\ when $R(\cL)=0$.
\end{proof}

The same holds for a classical generator $K$. Its column Gershgorin disks $D(K_{jj},|K_{jj}|)$ are contained in $D(-\nu_c,\nu_c)$ \cite{gershgorin}, so every nonzero eigenvalue of $K$ has negative real part and

\begin{equation}\label{eq:K}
R(K)=\max_{\lambda\in\spec(K)\setminus\{0\}}
\frac{|\lambda|^2}{-2\Re\lambda},
\end{equation}
satisfies $R(K) \leq \nu_c$, that is,
\begin{equation}
    \spec(K) \subseteq D(-R(K),R(K)) \subseteq D(-\nu_c,\nu_c) .
\end{equation}

We extend the notation by setting $R(\mathcal L)=+\infty$ if a nonzero eigenvalue lies on the imaginary axis. In that case no disk $D(-\nu,\nu)$ with finite $\nu$ contains $\spec(\cL)$, since then $|\lambda|^2+2\nu\Re\lambda=|\lambda|^2>0$.
This is the parameter introduced in equation (55) of \cite{corralling}. For $\lambda=-\Gamma+i\omega$, its contribution is $\Gamma/2+\omega^2/(2\Gamma)$. A bound on decay rates alone therefore does not determine a spectral disk.

The quantum counterpart of the maximal component is a pair of maximal singular vectors of an eigenoperator.

\begin{theorem}\label{thm:disk} Let $\Phi(X)=\sum_\alpha V_\alpha XV_\alpha^\dagger$ be completely positive, $G=\Phi^\dagger(I)=\sum_\alpha V_\alpha^\dagger V_\alpha$, and consider the GKLS generator
\begin{equation}\label{eq:positiveform}
\cL(X)=\Phi(X)-\frac12(GX+XG),
\end{equation}
i.e.\ with no Hamiltonian term in this representation. Then
\begin{equation}\label{eq:disk}
\spec(\cL)\subseteq D(-\nu,\nu),\qquad \nu=\|G\|_\infty.
\end{equation}
\end{theorem}
\begin{proof}
Clearly $G\ge0$. Let $Y\ne0$ satisfy $\cL^\dagger(Y)=\lambda Y$, so that $\lambda$ is an eigenvalue of $\cL^\dagger$ and $\bar\lambda$ one of $\cL$. Choose unit maximal singular vectors $u,v$ of $Y$, with $Yv=su$, $Y^\dagger u=sv$, $s=\|Y\|_\infty$. Define $g_u:=\langle u,Gu\rangle$ and $g_v:=\langle v,Gv\rangle$. One has
\begin{equation*}
    \langle u | \cL^\dagger(Y) v\rangle = \lambda \langle u | Y v\rangle = s \lambda ,
\end{equation*}
and hence 
\[  s\lambda = \langle u|\Phi^\dagger(Y)v\rangle  - \frac 12 \Big(  \langle u|GY v\rangle + \langle u|YGv\rangle \Big) =   \langle u|\Phi^\dagger(Y)v\rangle  - s\frac{g_u + g_v}{2} .\]
Using $\Phi^\dagger(Y) = \sum_\alpha V^\dagger_\alpha Y V_\alpha$ and the Cauchy--Schwarz inequality we have 
\begin{equation}\label{eq:CS}
|\langle u|\Phi^\dagger(Y)v\rangle| \le \| Y\|_\infty \sum_\alpha\|V_\alpha u\|\|V_\alpha v\|
\le s\sqrt{g_ug_v} .
\end{equation}
Hence we obtain the following bound for $\lambda$
 
\begin{equation}\label{eq:refined}
\left|\lambda+\frac{g_u+g_v}{2}\right|\le\sqrt{g_ug_v}.
\end{equation}
Now, write $a=(g_u+g_v)/2$ and $b=\sqrt{g_ug_v}$.
Since $0\le G\le\nu I$, arithmetic--geometric mean gives $0\le b\le a\le\nu$.
Thus
\[
|\lambda+\nu|\le|\lambda+a|+\nu-a\le b+\nu-a\le\nu ,
\]
which ends the proof since $\spec\mathcal L=\overline{\spec\mathcal L^\dagger}$ and the disk is conjugation invariant.
\end{proof}

Theorem~\ref{thm:disk} holds for every representation of the form \eqref{eq:positiveform}, i.e.\ with no Hamiltonian term in that representation; such representations are highly nonunique (Section~\ref{sec:optimization}). Its proof concerns eigenvalues only and does not use (complete) positivity of  $\id+\cL/\nu$, which may fail for every $\nu>0$ (amplitude damping, Example 5 of \cite{corralling}).

\begin{remark}[Single normal jump: an excluded lens]\label{rem:lens}
Let $g_-\le g_+$ be the extreme eigenvalues of $G$, so that $g_u,g_v\in[g_-,g_+]$ in \eqref{eq:refined}. For a single normal jump, $\cL=\D_A$ with $A=\operatorname{diag}(a_1,\dots,a_n)$ in a suitable basis, \eqref{eq:refined} becomes an equality: with $x_j=|a_j|^2\in[g_-,g_+]$, the eigenvalue $\lambda_{jk}=a_j\bar a_k-\frac12(x_j+x_k)$ of the matrix unit $E_{jk}$ satisfies
\[
\Bigl|\lambda_{jk}+\frac{x_j+x_k}{2}\Bigr|=\sqrt{x_jx_k}.
\]
Writing $\lambda_{jk}=-\Gamma+i\omega$, this reads $|\lambda_{jk}|^2=\Gamma(x_j+x_k)-\frac14(x_j-x_k)^2$, and minimizing over $x_j,x_k\in[g_-,g_+]$ gives $|\lambda_{jk}|^2\ge\min\{2g_-\Gamma,\ (g_-+g_+)\Gamma-\frac14(g_+-g_-)^2\}$. Hence no eigenvalue of $\D_A$ lies in the lens
\[
E=\Bigl\{z:|z+g_-|<g_-\Bigr\}\cap\Bigl\{z:\bigl|z+\tfrac{g_-+g_+}{2}\bigr|<\sqrt{g_-g_+}\Bigr\},
\]
which is nonempty if and only if $g_+<9g_-$. The region $D(-g_+,g_+)\setminus E$ is sharp: each of its nonzero points is an eigenvalue of $\D_A$ for some $A=\operatorname{diag}(a_1,a_2)$ with $|a_1|^2,|a_2|^2\in[g_-,g_+]$. For a unitary jump, $g_-=g_+=1$, the spectrum lies on the circle $|z+1|=1$. No such exclusion holds for several jumps. For $U=e^{-i\theta\sigma_3/2}$ and $r\in[0,1]$, the unital channel $\Phi(X)=\frac{1+r}{2}UXU^\dagger+\frac{1-r}{2}U\sigma_3X\sigma_3U^\dagger$ has $G=I$, and $\cL=\Phi-\id$ has the eigenvalues $re^{\pm i\theta}-1$; these fill the whole disk $D(-1,1)$. The lens $E$ contains the open disk $\operatorname{int}D(C_{\rm in},R_{\rm in})$ of Appendix~\ref{app:inside}, so for a single jump the spectrum lies in the ring $D(-g_+,g_+)\setminus\operatorname{int}D(C_{\rm in},R_{\rm in})$. Fig.~\ref{fig:counter}(b) shows such a ring for a qutrit.
\end{remark}

\begin{remark}\label{rem:positive}
Complete positivity is used only in the Cauchy--Schwarz step \eqref{eq:CS}. The same bound holds for every positive $\Phi$ (Lemma~\ref{lem:positive}), so Theorem~\ref{thm:disk} remains true for $\cL=\Phi-\frac12\{\Phi^\dagger(I),\cdot\}$ with $\Phi$ merely positive (Theorem~\ref{thm:diskpos} in Appendix~\ref{app:lemma}). Such generators define positive trace-preserving, but in general not completely positive, semigroups.
\end{remark}

\begin{remark}\label{rem:stadium}
The hypothesis of no Hamiltonian term cannot be dropped: if $H\notin\mathbb CI$, the generator $-i[H,\cdot]$ has nonzero purely imaginary eigenvalues. For $\cL=-i[H,\cdot]+\Phi-\frac12\{G,\cdot\}$ with $\Phi$ completely positive (or positive, Remark~\ref{rem:positive}), the term $i[H,Y]$ of $\cL^\dagger(Y)$ contributes $\langle u,i[H,Y]v\rangle=is(h_u-h_v)$, $h_u=\langle u,Hu\rangle$, so the proof gives $|\lambda-i(h_u-h_v)+\frac{g_u+g_v}{2}|\le\sqrt{g_ug_v}$ and
\begin{equation}\label{eq:stadium}
\spec(\cL)\subseteq D(-\nu,\nu)+i[-h,h],\qquad h=\lambda_{\max}(H)-\lambda_{\min}(H),
\end{equation}
a Minkowski sum with a vertical interval
of size $2h$ determined by the numerical range of the Hamiltonian, 
$W(H)=[\lambda_{min}, \lambda_{max}]$. For a Hermitian matrix $W(H)$ is the interval between its extreme eigenvalues \cite{HornJohnson2013}.

\end{remark}

For a jump representation $\{V_\alpha\}$ with arbitrary Hamiltonian one finds, by a direct computation of the superoperator trace,
\begin{equation}
     \Tr(-\cL)=n\tr G-\sum_\alpha|\tr V_\alpha|^2 .
\end{equation}
For the canonical representation, $\tr V_\alpha=0$, this gives $\Tr(-\cL)=n\tr G=n\sum_\alpha\gamma_\alpha$, which is the second equality in \eqref{eq:nu1}. Since $\Tr\cL$ is the sum of the $n^2$ eigenvalues, the spectrum is symmetric under complex conjugation and lies in the closed left half-plane,
\begin{equation}
\nu_{q,1}=\frac1n\sum_{k=1}^{n^2}(-\lambda_k)=\frac1n\sum_{k=1}^{n^2}|\Re\lambda_k| .
\end{equation}
So $\nu_{q,1}$ is intrinsic: it depends only on $\cL$, not on the chosen jumps, and the Hamiltonian does not enter it. Moreover, since $G\ge0$, for the canonical representation
\begin{equation}\label{eq:nu2nu1}
  \nu_{q,2}=\| G\|_\infty \leq \tr G = \frac 1n   \Tr(-\cL) = \nu_{q,1} .
\end{equation}
The escape rate is controlled by the operator $G=\Phi^\dagger(I)$ of every positive representation.

\begin{proposition}\label{prop:lower}
Let $\cL(X)=-i[H,X]+\Phi(X)-\frac12\{G,X\}$ with $\Phi$ positive, $G=\Phi^\dagger(I)$ and arbitrary Hermitian $H$. Then $0\le-\bra\psi\cL(P_\psi)\ket\psi\le\bra\psi G\ket\psi$ for every unit vector $\psi$, and hence $\nu_{q,3}(\cL)\le\|G\|_\infty$.
\end{proposition}
\begin{proof}
For $P_\psi=\ket\psi\bra\psi$, $\bra\psi[H,P_\psi]\ket\psi=0$ and $\bra\psi\{G,P_\psi\}\ket\psi=2\bra\psi G\ket\psi$. Positivity of $\Phi$ gives $0\le\bra\psi\Phi(P_\psi)\ket\psi\le\tr\Phi(P_\psi)=\bra\psi G\ket\psi$, and $-\bra\psi\cL(P_\psi)\ket\psi=\bra\psi G\ket\psi-\bra\psi\Phi(P_\psi)\ket\psi$.
\end{proof}

\begin{corollary}[Conjectures 1 and 2 of \cite{corralling}]\label{cor:conjectures}
For a GKLS generator $\cL\ne0$ whose canonical Hamiltonian vanishes,
\begin{equation}
\nu_{q,3}\le\nu_{q,2}\le\nu_{q,1},\qquad R(\cL)\le\nu_{q,2},
\qquad
\spec(\cL/\nu_{q,j})\subseteq D(-1,1),\quad j=1,2.
\end{equation}
\end{corollary}
\begin{proof}
The canonical representation has the form \eqref{eq:positiveform} with a completely positive jump map. Proposition~\ref{prop:lower} and \eqref{eq:nu2nu1} give the ordering, and Theorem~\ref{thm:disk} gives the inclusion for $j=2$, hence for $j=1$. Since $\cL\ne0$ forces $G\ne0$, the disk $D(-\nu_{q,2},\nu_{q,2})$ meets the imaginary axis only at $0$, and Proposition~\ref{prop:R} yields $R\le\nu_{q,2}$.
\end{proof}

The theorem applies to any specified representation \eqref{eq:positiveform}; canonical normalization is needed only to identify the particular intrinsic canonical parameters used in the earlier conjectures. A smaller representation-dependent value of $\nu$ need not contradict the canonical bound. For example, for $n=3$ and $V=\operatorname{diag}(1,0,0)$ one has $R=1/4$, the canonical values $\nu_{q,2}=4/9$ and $\nu_{q,1}=2/3$, while the representation $V-I/2$ gives $\|G\|_\infty=1/4=R$.
The canonical reading is essential. A generator written with $H=0$ in a non-canonical representation may have a nonzero canonical Hamiltonian; for $V=\sigma_-+3I$ the canonical Hamiltonian is $H_{\rm can}=-\frac32\sigma_y$, and one finds $\nu_{q,1}=1$ but $R=19/3$. The ordering $\nu_{q,2}\le\nu_{q,1}$ also fails if $\nu_{q,2}$ is evaluated in a non-canonical representation while $\nu_{q,1}$ keeps its canonical value.

%%%%%%%%%%%%%%%%%%%%%%%%%%%%%%%%%%%%%%%%%%%%%%%%%%%%%%%%%%%%%%%%%%%%%%%%%%%%%%%%%%
\section{Conjecture 3: the escape-rate disk}\label{sec:false}

\subsection{A counterexample}

For a GKLS generator and $P_\psi=\ket\psi\bra\psi$,
\begin{equation}\label{eq:variance}
-\tr[P_\psi\cL(P_\psi)]
=\sum_\alpha\left(\langle V_\alpha^\dagger V_\alpha\rangle_\psi
-|\langle V_\alpha\rangle_\psi|^2\right).
\end{equation}
The Hamiltonian contribution vanishes. For a diagonal jump $A=\operatorname{diag}(a_1,\dots,a_n)$ the dissipator $\D_A(X)=AXA^\dagger-\frac12\{A^\dagger A,X\}$ is diagonal in the matrix units,
\begin{equation}\label{eq:diagonaleigen}
\D_A(E_{jk})=\Bigl(a_j\bar a_k-\tfrac12\bigl(|a_j|^2+|a_k|^2\bigr)\Bigr)E_{jk}
=\Bigl(-\tfrac12|a_j-a_k|^2+i\,\Im(a_j\bar a_k)\Bigr)E_{jk}.
\end{equation}
The following example has zero canonical Hamiltonian, so the failure is not caused by omitting an explicit coherent term.

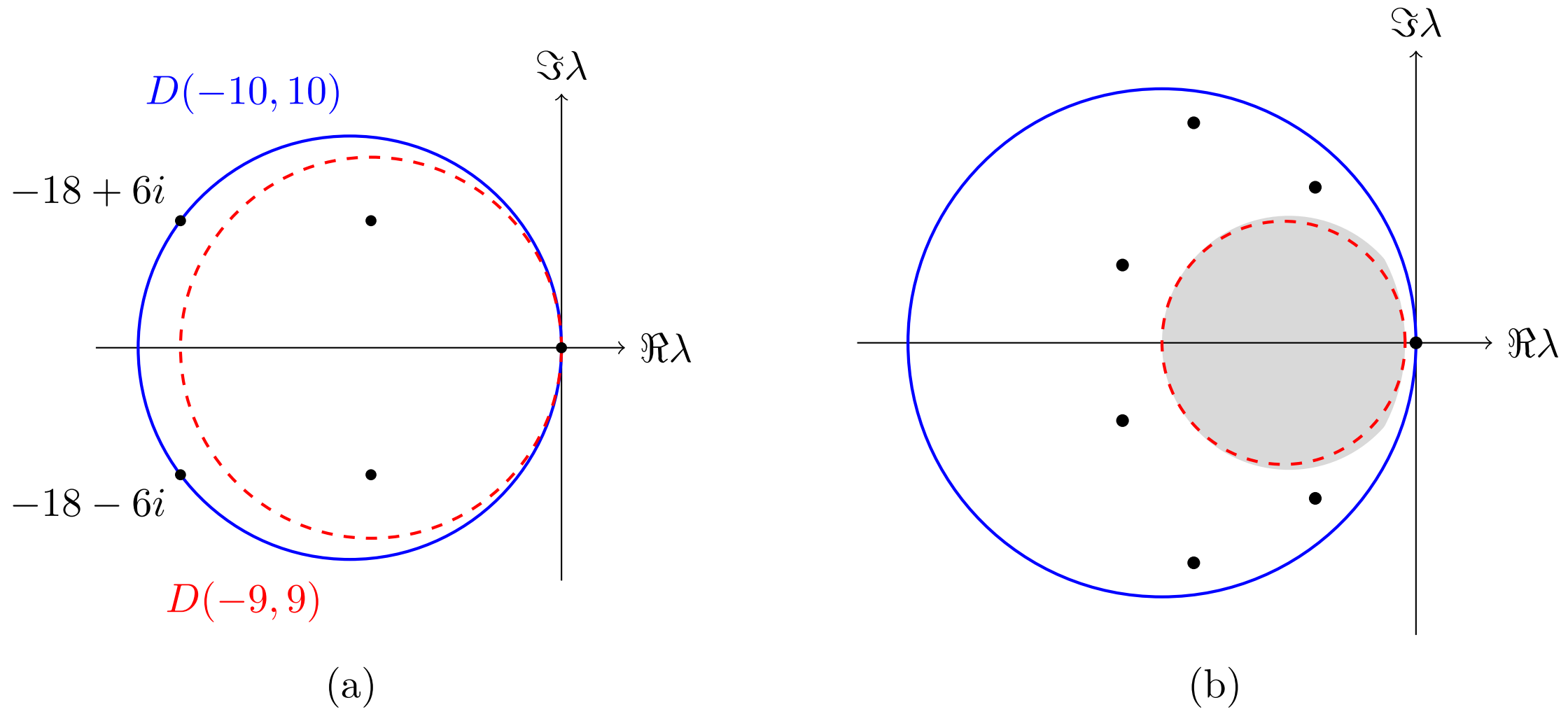
\begin{figure}[!htbp]
\centering
\begin{minipage}[t]{0.47\textwidth}
\centering
\begin{tikzpicture}[scale=.2]
\draw[->] (-22,0)--(3,0) node[right] {$\Re\lambda$};
\draw[->] (0,-11)--(0,12) node[above] {$\Im\lambda$};
\draw[blue,thick] (-10,0) circle (10);
\draw[red,dashed,thick] (-9,0) circle (9);
\foreach \x/\y in {-18/6,-18/-6,-9/6,-9/-6,0/0}{\fill (\x,\y) circle (.26);}
\node[above left] at (-18,6) {$-18+6i$};
\node[below left] at (-18,-6) {$-18-6i$};
\node[blue] at (-15,12) {$D(-10,10)$};
\node[red] at (-15,-12) {$D(-9,9)$};
\end{tikzpicture}\\[3pt]
(a)
\end{minipage}\hfill
\begin{minipage}[t]{0.47\textwidth}
\centering
\begin{tikzpicture}[scale=2.4]
\begin{scope}
\clip (-0.5,0) circle (0.5);
\fill[gray!30] (-0.75,0) circle (0.70711);
\end{scope}
\draw[->] (-2.2,0)--(0.3,0) node[right] {$\Re\lambda$};
\draw[->] (0,-1.15)--(0,1.15) node[above] {$\Im\lambda$};
\draw[blue,thick] (-1,0) circle (1);
\draw[red,dashed,thick] (-0.52145,0) circle (0.47855);
\foreach \x/\y in {-0.39645/0.61237,-0.39645/-0.61237,-0.875/0.86603,-0.875/-0.86603,-1.15533/0.30619,-1.15533/-0.30619,0/0}{\fill (\x,\y) circle (.025);}
\end{tikzpicture}\\[3pt]
(b)
\end{minipage}
\caption{(a) The qutrit counterexample. The eigenvalues $-18\pm6i$ lie on the boundary of the blue disk supplied by $\nu_{q,2}=10$ and outside the dashed disk proposed by $\nu_{q,3}=9$; the eigenvalues $-9\pm6i$ lie inside both. (b) A single qutrit, the qutrit example of Appendix~\ref{app:inside}, with $A=\operatorname{diag}(1,\tfrac{1}{\sqrt2}e^{i\pi/3},-\tfrac{\sqrt3}{2}i)$. Blue: the outer disk $D(-g_{\max},g_{\max})=D(-1,1)$. Grey: the open excluded lens $E$ of Remark~\ref{rem:lens}. Red dashed: the boundary of a smaller eigenvalue-free open disk, $\operatorname{int}D(C_{\rm in},R_{\rm in})$. Dots: the six nonzero eigenvalues and $0$.
{All eigenvalues lie inside the blue disk but outside the grey excluded lens \(E\). The red dashed circle illustrates the weaker inner-disk bound of Appendix C.}
\label{fig:counter} }
\end{figure}

\begin{example}[A qutrit counterexample \cite{Zhang}]\label{ex:counter}
Let
\begin{equation}\label{eq:A}
A=\operatorname{diag}(1+3i,1-3i,-2),\qquad
\cL=\D_A,\quad \cL(X)=AXA^\dagger-\tfrac12\{A^\dagger A,X\}.
\end{equation}
The jump is traceless, normal and satisfies $A^\dagger A=\operatorname{diag}(10,10,4)$. Since $A$ is diagonal, the escape rate depends only on the moduli of the components of $\psi$; for $|\psi_1|^2=p$, $|\psi_2|^2=q$, $|\psi_3|^2=r$, with $p+q+r=1$, equation~\eqref{eq:variance} gives
\begin{align}
\langle A^\dagger A\rangle-|\langle A\rangle|^2
%=10-6r-(1-3r)^2-9(p-q)^2 = 
=9[1-r^2-(p-q)^2]\le9.
\end{align}
Equality holds for $p=q=1/2$, $r=0$. By \eqref{eq:diagonaleigen},
\[
\cL(E_{12})= (-18+6i) E_{12} , \quad \cL(E_{13})= (-9-6i) E_{13} , \quad \cL(E_{23})= (-9+6i) E_{23} ,
\]
the modes $E_{kj}$ carry the complex conjugate eigenvalues, and $\cL(E_{jj})=0$. Hence $\spec\cL=\{0^{(3)},-18\pm6i,(-9\pm6i)^{(2)}\}$ and $R(\cL)=\max\{360/36,117/18\}=10$. Finally
\begin{equation}
\nu_{q,3}=9,\qquad R(\cL)=\nu_{q,2}=10,\qquad \nu_{q,1}=24.
\end{equation}
In particular $\spec(\cL/\nu_{q,3})\not\subseteq D(-1,1)$, see Fig.~\ref{fig:counter}(a). The disk of Theorem~\ref{thm:disk} is saturated, $|-18+6i+10|=10$. The generator is unital and normal as a superoperator, so neither unitality nor normality rescues Conjecture~3.
\end{example}

The example is not an isolated accident. Appendix~\ref{sec:geometry} analyzes the family $A_b=\operatorname{diag}(1+ib,1-ib,-2)$, which contains \eqref{eq:A} at $b=3$: Conjecture~3 fails for every $b\ne0$ except $b^2=3$, the ratio $R/\nu_{q,3}$ is maximal, equal to $9/8$, at $b^2=7$, and for $b^2\ge3$ the disk of Theorem~\ref{thm:disk} is saturated.
Fig.~\ref{fig:counter}(a) shows the spectrum of \eqref{eq:A} against the two disks, and Fig.~\ref{fig:counter}(b) a single-jump qutrit whose spectrum avoids the inner disk of Appendix~\ref{app:inside}, so that it lies in the ring \eqref{eq:ring1}.

\vspace{.1cm}

{

%\subsection{A new weaker bound: $\spec \cL \in D(-4 \nu_{q,3},4\nu_{q,3})$}

Although Conjecture~3 fails in general, replacing $\nu_{q,3}$ by $4\nu_{q,3}$
yields a valid disk bound in every finite dimension, provided the canonical Hamiltonian vanishes.  

\begin{lemma}\label{lem:four}
Let $V_1,\ldots,V_m\in M_n(\mathbb C)$ be traceless, and let
\[
\nu_{q,3} :=\max_{\|\psi\|=1}\sum_\alpha
\big( \|V_\alpha\psi\|^2-|\langle\psi,V_\alpha\psi\rangle|^2\big).
\]
Then, $G= \sum_\alpha V_\alpha^\dagger V_\alpha$ satisfies 
\begin{equation}\label{eq:four}
\left\| G \right\|_\infty\leq4\nu_{q,3}.
\end{equation}
\end{lemma}
For the proof see Appendix \ref{App-4nu}.

\begin{theorem}\label{thm:sharp4}
For every GKLS generator with vanishing canonical Hamiltonian,
\begin{equation}\label{eq:sharp4}
R(\cL)\leq\nu_{q,2} \leq 4\nu_{q,3} .
\end{equation}
Thus 
$
\spec(\cL) \subseteq D(-4\nu_{q,3},4\nu_{q,3}).
$
The constant $4$ is optimal as a single constant valid in all finite dimensions.
\end{theorem}
\begin{proof}
The canonical jumps are traceless, so the second inequality is
Lemma~\ref{lem:four}. The first follows from Theorem~\ref{thm:disk}.
To establish sharpness, take $n\geq4$, $b>0$, and the traceless normal jump

\begin{equation}\label{eq:Anb}
A_{n,b}=\operatorname{diag}\left(1+ib,1-ib,
\underbrace{-\frac{2}{n-2},\ldots,-\frac{2}{n-2}}_{n-2\text{ entries}}\right),
\qquad \cL_{n,b}=\mathcal{D}_{A_{n,b}}.
\end{equation}
The same diagonal expression at $n=3$ and $b=3$ recovers \eqref{eq:A};
the identities below are used for $n\ge4$. Put
$h_n=n/(n-2)$. The coherence $E_{12}$ has eigenvalue
\[
\lambda_{12}=-2b^2+2ib,
\qquad \frac{|\lambda_{12}|^2}{-2\Re\lambda_{12}}=1+b^2.
\]
Because $n\geq4$, $\nu_{q,2} =1+b^2$, and the disk theorem gives
\begin{equation}\label{eq:Rfamily}
R(\cL_{n,b})=\nu_{q,2}=1+b^2.
\end{equation}
Let $\psi=(\psi_1,\ldots,\psi_n)^T$ and $p=|\psi_1|^2$, $q=|\psi_2|^2$, $u=p+q$.
All remaining diagonal entries coincide, so the variance is exactly
\[
e(\psi)=b^2u-b^2(p-q)^2+h_n^2u(1-u).
\]
For fixed $u$, the maximum is attained at $p=q=u/2$. If
$0<b^2\leq h_n^2$, maximizing $e(\psi)$  over $0\leq u\leq1$
gives
\begin{equation}\label{eq:escapeFamily}
\nu_{q,3}=\frac{(h_n^2+b^2)^2}{4h_n^2},\qquad
\frac{R}{\nu_{q,3}}=\frac{4h_n^2(1+b^2)}{(h_n^2+b^2)^2}.
\end{equation}
If $b^2\geq h_n^2$, the escape rate is $b^2$. Taking $b=1/n$ in
\eqref{eq:escapeFamily} gives $R/\nu_{q,3}\to4$ as $n\to\infty$.
Thus every constant strictly below $4$ fails in some finite dimension.
Here $b=1/n$ is strictly positive for every $n$; the mode $\lambda_{12}$ must not be
set equal to zero before evaluating its contribution to $R$.
\end{proof}

}

\FloatBarrier

\subsection{Conjecture 3 holds for qubits}\label{sec:qubit}

Conjecture 3 holds for qubits without a canonical Hamiltonian. Every purely dissipative qubit generator in canonical form can be written as

\begin{equation}\label{eq:qubitgen}
\cL(\rho)=\sum_{j,k=1}^3C_{jk}\Bigl(\sigma_j\rho\sigma_k-\tfrac12\{\sigma_k\sigma_j,\rho\}\Bigr),\qquad C=C^\dagger\ge0,
\end{equation}
with the Kossakowski matrix $C$ taken in the (unnormalized) Pauli basis. Write $C=C_R+iC_I$ with $C_R$ real symmetric and $C_I$ real antisymmetric. Since $\overline C=C^{\rm T}\ge0$, also $C_R=\frac12(C+\overline C)\ge0$; let $c_1\ge c_2\ge c_3\ge0$ be its eigenvalues. Define a Bloch vector $\bm{r} =(x_1,x_2,x_3)$, with $x_k = {\rm Tr}(\rho \sigma_k)$. 

\begin{theorem}[Conjecture 3 for qubits]\label{thm:qubit}
For every qubit GKLS generator (\ref{eq:qubitgen}):

\begin{enumerate}
\item[(i)] in Bloch coordinates $\rho=\frac12(I+r\cdot\sigma)$ the dynamics is $\dot {\bm r}=A{\bm r} + {\bm b}$ with
\begin{equation}\label{eq:bloch}
A=-2\bigl(\tr(C_R)\,I_3-C_R\bigr),\qquad b_l=-2\sum_{j,k}\varepsilon_{jkl}(C_I)_{jk};
\end{equation}
\item[(ii)] the spectrum is real, $\spec\cL=\{0,\,-2(c_2+c_3),\,-2(c_3+c_1),\,-2(c_1+c_2)\}$, and $R(\cL)=c_1+c_2$;
\item[(iii)] $R(\cL)\le\nu_{q,3}(\cL)$, so $\spec(\cL/\nu_{q,3})\subseteq D(-1,1)$ whenever $\cL\neq0$;
\item[(iv)] if $\cL$ is unital, then $\nu_{q,3}=R=c_1+c_2$ and $\nu_{q,2}=c_1+c_2+c_3$ (the optimized representation scales are given in Example~\ref{sec:pauli}).
\end{enumerate}
\end{theorem}
\begin{proof}
(i) Using the following identity
%emph{The linear and affine parts of the Bloch equation.} With 

$$ \tr(\sigma_a\sigma_b\sigma_c\sigma_d)=2(\delta_{ab}\delta_{cd}-\delta_{ac}\delta_{bd}+\delta_{ad}\delta_{bc}) $$ 
one finds 

$$\frac12\tr\bigl(\sigma_l\,\cL(\sigma_m)\bigr)=\sum_{jk}C_{jk}(\delta_{lj}\delta_{mk}+\delta_{lk}\delta_{jm}-2\delta_{lm}\delta_{jk})=2\Re C_{lm}-2\delta_{lm}\tr C ,$$ 
which is the first formula. Moreover 

$$\cL(I)=\sum_{jk}C_{jk}[\sigma_j,\sigma_k]=2i\sum_{jkl}C_{jk}\varepsilon_{jkl}\sigma_l ,$$
the symmetric part $C_R$ drops out and $\cL(I/2)=\frac12 {\bm b}\cdot{\bm \sigma}$ gives the second formula.

(ii) %\emph{Why the drift does not affect the eigenvalues.} 
In the basis $(I,\sigma_1,\sigma_2,\sigma_3)$ the matrix of $\cL$ is block lower triangular with diagonal blocks $0$ and $A$

\[ \begin{pmatrix}0 & {\bm 0}^{\rm T}\\ {\bm b} & A\end{pmatrix}\]
so $\spec\cL=\{0\}\cup\spec A$. The matrix $A$ is real symmetric with eigenvalues $-2(\tr C_R-c_l)\le0$. For a nonzero real eigenvalue $\lambda<0$, $|\lambda|^2/(-2\Re\lambda)=-\lambda/2$, and \eqref{eq:R} gives $R=\tr C_R-c_3=c_1+c_2$.

(iii) %\emph{Antipodal pure states remove the drift.} 
For a pure state with Bloch vector ${\bm n}$, $|{\bm n}|=1$, we have $\cL(P_n)=\frac12(A {\bm n} + {\bm b})\cdot{\bm \sigma}$ and therefore
\begin{equation}\label{eq:qubitescape}
e({\bm n}):=-\tr[P_n\cL(P_n)]=-\tfrac12\,{\bm n}\cdot(A{\bm n} +{\bm b})=\tr C_R-{\bm n}^{\rm T}C_R {\bm n}-\tfrac12\,{\bm b}\cdot {\bm n} .
\end{equation}
Let ${\bm u}_3$ be a unit eigenvector of $C_R$ for $c_3$. Then $e({\bm u}_3)+e(-{\bm u}_3)=2(\tr C_R-c_3)=2R$, so $\nu_{q,3}\ge\max\{e({\bm u}_3),e(-{\bm u}_3)\}\ge R$. The disk inclusion follows from Proposition~\ref{prop:R} and the nesting of the disks, since $\nu_{q,3}\ge R$, and $\nu_{q,3}\ge R=c_1+c_2\ge c_1>0$ for $\cL\ne0$: indeed $\cL\neq0$ forces $C\ne0$, hence $\tr C_R=\tr C>0$.

(iv) %\emph{Reduction of the unital case to Pauli generators.} 
Unitality means ${\bm b=0}$, i.e.\ $C_I=0$. Then $e({\bm n})=\tr C_R-{\bm n}^{\rm T}C_R{\bm n}$, whose maximum on the sphere is $\tr C_R-c_3=R$. A real symmetric $C$ is diagonalized by a rotation $O\in SO(3)$, and rotations of the Bloch sphere are implemented by unitary conjugations, which do not change any of the scales. In the rotated frame $\cL$ is the Pauli generator \eqref{eq:pauligen} with $\gamma_j=c_j$, whose canonical representation has $G=(c_1+c_2+c_3)I$. 
\end{proof}

The proof shows where the argument breaks in higher dimensions and in the presence of a Hamiltonian. A canonical Hamiltonian $H=h\cdot\sigma/2$ adds to $A$ the antisymmetric matrix $r\mapsto h\times r$, which in general produces complex eigenvalues while leaving $\nu_{q,3}$ unchanged. Without a Hamiltonian the qubit Bloch matrix is always symmetric, whereas already for a qutrit the generator \eqref{eq:A} has complex eigenvalues produced by a single normal jump.

\begin{proposition}[Equality for qubit generators]\label{prop:qubitequality}
For a qubit GKLS generator with vanishing canonical Hamiltonian,
$R=\nu_{q,3}$ if and only if the generator is unital.
\end{proposition}
\begin{proof}
The unital implication is Theorem~\ref{thm:qubit}(iv).
Conversely, assume $\nu_{q,3}=R$ and put $B=C_R-c_3I\ge0$.
Equation~\eqref{eq:qubitescape} says that
${\bm n}^{\rm T}B {\bm n}+\tfrac12 {\bm b}\cdot {\bm n}\ge0$ for all unit vectors ${\bm n}$.
Choose a unit ${\bm u}\in\ker B$. Testing at ${\bm u}$ and $-{\bm u}$ gives ${\bm b}\cdot {\bm u}=0$.
For any unit ${\bm v}\perp {\bm u}$, set ${\bm n}(\theta)=\cos\theta\,{\bm u}+\sin\theta\,{\bm v}$.
Since $B{\bm u}=0$, the same inequality becomes
\[
\sin^2\theta\,{\bm v}^{\rm T}B {\bm v} + \tfrac12\sin\theta\,{\bm b}\cdot {\bm v}\ge0.
\]
Taking $\theta\to0$ from both sides forces ${\bm b}\cdot {\bm v}=0$.
Thus ${\bm b}=0$, which is unitality. The argument also covers degenerate $c_3$ and the zero generator.
\end{proof}
For amplitude damping, for example, the inequality is strict: $R=\gamma/2$ and $\nu_{q,3}=\gamma$.

\begin{remark}\label{rem:qubitPTP}
The Bloch formula in (i) remains valid for Hermitian $C$ without $C\ge0$. If $\cL$ is any qubit generator of a positive trace-preserving semigroup with vanishing canonical Hamiltonian, then $A$ is symmetric. Positivity of the semigroup makes the Bloch ball invariant, so
${\bm n}\cdot(A{\bm n}+{\bm b})\le0$ for every unit ${\bm n}$. Adding the inequalities at ${\bm n}$ and $-{\bm n}$ gives ${\bm n}^{\rm T}A{\bm n}\le0$,
and therefore $A\le0$. For a unit eigenvector ${\bm u}$ of $A$ belonging to $\lambda_{\min}(A)$ one has $e({\bm u})+e(-{\bm u})=-\lambda_{\min}(A)=2R(\cL)$. Hence $R\le\nu_{q,3}$ for all such generators, and Conjecture~3 holds for every qubit positive trace-preserving (PTP) generator without a canonical Hamiltonian.

\end{remark}

%%%%%%%%%%%%%%%%%%%%%%%%%%%%%%%%%%%%%%%%%%%%%%%%%%%%%%%%%%%%%%%%%%%%%%%%%%%%%%%%%%%%%%%
% Replacement for the former subsections 3.3 and 3.4.
% Uses the manuscript macros \cL, \tr, \spec, and its theorem environments.
% Cross-references and bibliography keys refer to the complete manuscript.

\subsection{Universal decay-rate bound for positive trace-preserving semigroups}
\label{sec:universal-decay}

{

The escape rate controls the real parts of all generator eigenvalues, even when
it does not determine a disk containing the entire spectrum. The following
result holds for every positive trace-preserving semigroup in finite dimension;
unitality, complete positivity, and the absence of a Hamiltonian term are not
required. We write for any rank-1 projector $P$
\[
e(P)=-\tr[P\cL(P)],\qquad
\nu_{q,3}=\max_{P}e(P).   %\text{ rank-one projection}}
\]
The nonnegativity of $e(P)$, and hence the agreement with
\eqref{eq:nu3}, follows from the first step of the proof below.

\begin{theorem}[Universal decay-rate bound]\label{thm:universal-decay-rate}
Let $\cL:M_n(\mathbb C)\to M_n(\mathbb C)$ generate a positive
trace-preserving semigroup $e^{t\cL}$, $t\ge 0$. 
Then, with no assumptions of unitality, complete positivity, or
absence of a Hamiltonian term, one has the following bound for the maximal relaxation rate of $\cL$
\begin{equation}\label{eq:universal-decay-rate}
\Gamma_{\max}
:=\max_{\lambda\in\spec\cL}(-\Re\lambda)
\le 2\nu_{q,3}(\cL).
\end{equation}
The factor $2$ is optimal.
\end{theorem}

\begin{proof}
The generator $\cL$ preserves Hermiticity and annihilates the
trace. Fix an eigenvalue $\lambda=-\Gamma+i\omega$ and a nonzero
eigenoperator $Z$, so that $\cL(Z)=\lambda Z$.
Multiplying $Z$ by a phase leaves its eigenvalue equation
unchanged. Choose a phase maximizing $
\left\| e^{i\theta}Z+e^{-i\theta}Z^\dagger \right\|_1$. 
After absorbing this phase into $Z$, write $Z=X+iY$, where
$X=X^\dagger$, $Y=Y^\dagger$, and $X\ne0$. Our choice implies
\begin{equation}\label{eq:phase-max-36}
\|\cos\theta\,X-\sin\theta\,Y\|_1\le\|X\|_1
\qquad (\theta\in\mathbb R).
\end{equation}
Let $S=\operatorname{sgn}(X)$, and we set $S=0$ on $\ker X$.
Then $S=S^\dagger$, $\|S\|_\infty\le1$, and
$\tr(SX)=\|X\|_1$. By trace-norm duality and
\eqref{eq:phase-max-36}, the smooth real-valued function
\[
g(\theta)=\tr\bigl[S(\cos\theta\,X-\sin\theta\,Y)\bigr]
\]
satisfies $g(\theta)\le\|X\|_1=g(0)$. Hence a smooth function $g(\theta)$ has a maximum at $\theta=0$ which implies

\begin{equation}\label{eq:phase-orth-36}
0=g'(0)=-\tr(SY).
\end{equation}
Taking Hermitian parts of $\cL(Z)=(-\Gamma+i\omega)Z$,
we obtain
\[
\cL(X)=-\Gamma X-\omega Y.
\]
Consequently, \eqref{eq:phase-orth-36} gives the identity
\begin{equation}\label{eq:spectral-identity-36}
-\tr[S\cL(X)]
=\Gamma\tr(SX)+\omega\tr(SY)
=\Gamma\|X\|_1.
\end{equation}
In particular, we did not assume that $X$ is itself an
eigenoperator when $\omega\ne0$.
Now, choose an orthonormal eigenbasis of $X$ and write
\[
X=\sum_{j=1}^{n}x_jP_j,\qquad
S=\sum_{j=1}^{n}s_jP_j,\qquad
s_j=\operatorname{sgn}(x_j),\quad x_j\in\mathbb R,
\]
with rank-one orthogonal projections $P_j$.
Set
\[
K_{ij}:=\tr[P_i\cL(P_j)].
\]
 $K$ is a Kolmogorov generator in the column convention: $K_{ij} \geq 0$ for $i \neq j$ and $\sum_i K_{ij}=0$.  Its exit rates are
\begin{equation}\label{eq:exit-36}
r_j:=-K_{jj}=\sum_{i\ne j}K_{ij}
=-\tr[P_j\cL(P_j)]
\in[0,\nu_{q,3}].
\end{equation}
It follows that
\begin{align}
-\tr[S\cL(X)]
&=-\sum_{i,j}s_iK_{ij}x_j =\sum_j\sum_{i\ne j}K_{ij}
   \bigl(|x_j|-s_ix_j\bigr) \le 2\sum_j\sum_{i\ne j}K_{ij}|x_j| \notag\\
& =2\sum_j r_j|x_j|
\le\, 2\nu_{q,3}\sum_j|x_j| =\, 2\nu_{q,3}\|X\|_1.
\label{eq:classical-bound-36}
\end{align}
Here we used only $|s_i|\le1$, $K_{ij}\ge0$ for
$i\ne j$, and \eqref{eq:exit-36}. Combining
\eqref{eq:spectral-identity-36} and
\eqref{eq:classical-bound-36}, and dividing by
$\|X\|_1>0$, yields
$-\Re\lambda=\Gamma\le2\nu_{q,3}$.
Since $\lambda$ was arbitrary, this proves the assertion.

For optimality, the qubit dephasing generator
$\cL(\rho)=\gamma(\sigma_3\rho\sigma_3-\rho)$, $\gamma>0$, has
$\nu_{q,3}=\gamma$ and
$\spec\cL=\{0,0,-2\gamma,-2\gamma\}$. Therefore
$\Gamma_{\max}=2\nu_{q,3}$.
\end{proof}

\begin{corollary}[Real spectrum]\label{cor:real-spectrum}
Let $\cL$ generate a positive trace-preserving semigroup on $M_n$.
Every real eigenvalue belongs to $[-2\nu_{q,3},0]$.
If the entire spectrum is real, then
\begin{equation}\label{eq:real-spectrum-disk}
\spec\cL\subseteq[-2\nu_{q,3},0]
\subseteq D(-\nu_{q,3},\nu_{q,3}),
\end{equation}
and $R(\cL)\le\nu_{q,3}$.
\end{corollary}

In particular, Corollary~\ref{cor:real-spectrum} applies to
Hilbert--Schmidt self-adjoint generators. The same conclusion holds
when the entire generator is self-adjoint in a weighted inner product,
as in the following detailed-balance setting
\cite{Alicki1976,KFGV1977,FagnolaUmanita2007,FagnolaUmanita2010,CarlenMaas2020}.

\begin{corollary}[Detailed balance]\label{cor:db}
Let $\cL$ generate a positive trace-preserving semigroup on $M_n$, and
let $\sigma>0$ be a faithful density matrix. Suppose that the
\emph{entire} Heisenberg generator $\cL^\dagger$, including any
Hamiltonian part, is self-adjoint with respect to either inner product
\begin{align*}
\langle X,Y\rangle_{\sigma,\mathrm{GNS}}
&=\tr(\sigma X^\dagger Y),\\
\langle X,Y\rangle_{\sigma,\mathrm{KMS}}
&=\tr(\sigma^{1/2}X^\dagger\sigma^{1/2}Y).
\end{align*}
Then $\cL(\sigma)=0$ and
\[
\spec\cL\subseteq[-2\nu_{q,3},0],\qquad
R(\cL)\le\nu_{q,3}.
\]
In particular, the escape-rate disk inclusion holds.
\end{corollary}
\begin{proof}
Faithfulness makes either inner product positive definite.
Self-adjointness implies that $\cL^\dagger$ has real spectrum, and
so does its Hilbert--Schmidt adjoint $\cL$.
Corollary~\ref{cor:real-spectrum} therefore applies.
Moreover, trace preservation gives $\cL^\dagger(I)=0$. For either
of the displayed inner products,
\[
\tr[\sigma\cL^\dagger(X)]
=\langle I,\cL^\dagger(X)\rangle_{\sigma,i}
=\langle\cL^\dagger(I),X\rangle_{\sigma,i}=0
\]
for all $X$ and $i\in \{\mbox{GNS,KMS}\}$ which is equivalent to $\cL(\sigma)=0$.
\end{proof}
Here self-adjointness of the dissipative part alone is not sufficient:
the hypothesis concerns the entire generator, not a detailed-balance
condition imposed only after subtracting a Hamiltonian derivation.

\begin{remark}[Decay-rate control does not imply an escape-rate disk]
\label{rem:decay-vs-disk}
For $\lambda=-\Gamma+i\omega$ and $\nu=\nu_{q,3}$,
\[
|\lambda+\nu|\le\nu
\quad\Longleftrightarrow\quad
\omega^2\le\Gamma(2\nu-\Gamma).
\]
The universal estimate $0\le\Gamma\le2\nu$ does not impose this
additional restriction on $\omega$. In particular, the qutrit generator with vanishing canonical
Hamiltonian of Example~\ref{ex:counter} has
$\nu=9$ and eigenvalues $-18\pm6i$. It saturates the decay-rate bound,
$\Gamma_{\max}=18=2\nu$, but
\[
|-18+6i+9|=\sqrt{117}>9.
\]
Thus its spectrum is not contained in $D(-9,9)$, despite satisfying
the universal decay-rate bound. The generator is both unital and normal,
so neither property repairs the disk inclusion. More generally, every
violation of the escape-rate disk must involve a nonreal eigenvalue;
real eigenvalues are covered by Corollary~\ref{cor:real-spectrum}.
\end{remark}

\begin{remark}[Comparison with the total canonical rate]
For GKLS generators, the relaxation-rate bound
\cite{Relax1,Relax2} gives $\Gamma_{\max}\le\nu_{q,1}$.
Combining it with Theorem~\ref{thm:universal-decay-rate} yields
\[
\Gamma_{\max}
\le\min\{\nu_{q,1},2\nu_{q,3}\}.
\]
If the spectrum is real, it is consequently contained in both
$D(-\nu_{q,1}/2,\nu_{q,1}/2)$ and $D(-\nu_{q,3},\nu_{q,3})$.
\end{remark}

}

\section{Optimizing the representation and the spectral disk}\label{sec:optimization}

Representation \eqref{eq:positiveform} is not unique and we write $\nu_2(\Phi)=\|\Phi^\dagger(I)\|_\infty$. The canonical value is $\nu_{q,2}=\nu_2(\Phi_{\rm can})$, where the canonical $\Phi_{\rm can}$ is defined in terms of traceless Kraus operators $V_\alpha$. The canonical jump map is unique (its Kraus operators only up to a unitary mixing). If a map $\Phi$ realizes \eqref{eq:positiveform}, then for $c\ge0$ the map $\Phi+c\id$ realizes the same generator with $G$ replaced by $G+cI$. Consequently
\begin{equation}\label{eq:shift}
\nu_2(\Phi+c\id)=\nu_2(\Phi)+c.
\end{equation}
The added jump $\sqrt c I$ has zero dissipator. Thus a jump count, and its maximum intensity, can change without changing the dynamics. Canonical representation  defines $\nu_{q,2}$ intrinsically, but does not make it the smallest representation scale.

For a Hermitian jump $A$, replacing it by $A-cI$ with real $c$ leaves the dissipator unchanged. Taking
\begin{equation}\label{eq:shiftexample}
A=\operatorname{diag}(-2,-1,3)
\end{equation}
gives a canonical scale $\nu_{q,2}=9$, while $A-I/2$ gives $\nu_2(\Phi)= 25/4$. The eigenvalue associated with the extreme levels is $-25/2$, so the latter scale is optimal even spectrally.

For a fixed Hermiticity-preserving, trace-annihilating map $\cL$, define

\begin{equation}\label{eq:nucp}
\nu_{\CP}(\cL):=\inf\left\{\|G\|_\infty:
\Phi_G=\cL+\tfrac12\{G,\cdot\}\text{ is CP},\ G=G^\dagger\right\},
\end{equation}
with value $+\infty$ if the feasible set is empty. No Hamiltonian term is subtracted during this optimization: the entire generator $\cL$ is kept fixed. The normalization of $G$ is automatic: for every $X$,
$\tr\Phi_G(X)=\tr\mathcal L(X)+\tfrac12\tr(GX+XG)=\tr(GX)$.
Hence $\Phi_G^\dagger(I)=G$ and complete positivity of $\Phi_G$ then forces $G\ge0$.

Define a Choi matrix $C(\Psi)=\sum_{ij}E_{ij}\otimes\Psi(E_{ij})$ and the unnormalized vector $|\Gamma\rangle=\sum_i\ket{ii}$ together with  $P_\Gamma = | \Gamma\rangle \langle \Gamma|$. By the Choi criterion \cite{choi}, $\Phi_G$ is CP if and only if
\begin{equation}
C(\Phi_G)=C(\cL)+\frac12\left[(I\otimes G)P_\Gamma+P_\Gamma(I\otimes G)\right]\ge0.
\end{equation}
Thus \eqref{eq:nucp} formulates precisely the following semidefinite program (SDP) \cite{vandenberghe}
\begin{equation}\label{eq:sdp}
\begin{aligned}
\text{minimize}\quad &t\\
\text{over}\quad &G=G^\dagger,\quad t\in\mathbb R,\\
\text{subject to}\quad &0\le G\le tI,\\
&C(\cL)+\tfrac12[(I\otimes G)P_\Gamma+P_\Gamma(I\otimes G)]\ge0.
\end{aligned}
\end{equation}

\begin{proposition}\label{prop:attainment}
Whenever \eqref{eq:sdp} is feasible, its minimum is attained. Moreover
\begin{equation}
\max\{R(\cL),\nu_{q,3}\}\le\nu_{\CP}(\cL).
\end{equation}
If $\cL$ is a GKLS generator whose canonical Hamiltonian vanishes, then
\begin{equation}\label{eq:cpchain}
\max\{R(\cL),\nu_{q,3}\}\le\nu_{\CP}(\cL)\le\nu_{q,2}\le\nu_{q,1}.
\end{equation}
\end{proposition}
\begin{proof}
A feasible pair $(G_0,t_0)$ allows the restriction $0\le t\le t_0$. The remaining feasible set is closed and bounded because $0\le G\le tI$, hence compact. For every feasible $G$ Theorem~\ref{thm:disk} gives $\spec\cL\subseteq D(-\|G\|_\infty,\|G\|_\infty)$; if $G\ne0$ this disk meets the imaginary axis only at $0$, so Proposition~\ref{prop:R} yields $R(\cL)\le\|G\|_\infty$, and if $G=0$ then $\Phi_0^\dagger(I)=0$ forces $\Phi_0=0$ and $\cL=0$. Proposition~\ref{prop:lower} gives $\nu_{q,3}\le\|G\|_\infty$ for every feasible $G$. When the program is infeasible the inequality $R\le\nu_\CP=+\infty$ is void. The canonical CP jump map is feasible when its Hamiltonian vanishes, and the last inequality follows from Corollary~\ref{cor:conjectures}.
\end{proof}
If $C(\Phi_G)=\sum_\alpha|v_\alpha\rangle\langle v_\alpha|$ at an optimum, write
$v_\alpha=(I\otimes V_\alpha)\Gamma$. The matrices $V_\alpha$ give a minimizing Kraus representation,
and $\sum_\alpha V_\alpha^\dagger V_\alpha=G$ follows from trace annihilation. The program does not require a priori selection of the number or shape of the jumps.

\begin{example}[Hermitian single-jump generators]\label{sec:hermitian}
Let $A=A^\dagger$, with extreme eigenvalues $a_-$ and $a_+$. The generator $\D_A(X)=AXA-\frac12\{A^2,X\}$ has eigenvalues $-(a_i-a_j)^2/2$, so $R(\D_A)=(a_+-a_-)^2/4$. The shifted jump $A-(a_++a_-)/2$ gives the same dissipator and the feasible point $G=(A-(a_++a_-)/2)^2$ of \eqref{eq:sdp} with $\|G\|_\infty=(a_+-a_-)^2/4$. Together with Proposition~\ref{prop:attainment},
\begin{equation}\label{eq:hermitian}
R(\D_A)=\nu_{\CP}(\D_A)=\frac{(a_+-a_-)^2}{4}.
\end{equation}
The pure-state escape rate equals the same number: it is the maximum variance of $A$. Every distribution supported in $[a_-,a_+]$ has variance at most $(a_+-a_-)^2/4$, %(the Popoviciu inequality \cite{popoviciu}), 
and the bound is attained by an equal superposition of the extreme eigenvectors. This explains the success of Conjecture 3 for Hermitian single jumps without extending it to general normal jumps.

\end{example}

\begin{example}[Pauli generators]\label{sec:pauli}
  
For
\begin{equation}\label{eq:pauligen}
\cL(X)=\sum_{j=1}^3\gamma_j(\sigma_jX\sigma_j-X),\qquad\gamma_1\ge\gamma_2\ge\gamma_3\ge0,
\end{equation}
the eigenvalues on $\sigma_1,\sigma_2,\sigma_3$ are $-2(\gamma_2+\gamma_3)$, $-2(\gamma_3+\gamma_1)$ and $-2(\gamma_1+\gamma_2)$; some of them may vanish. Thus
\begin{equation}
R=\gamma_1+\gamma_2,
\qquad \nu_{q,2}=\gamma_1+\gamma_2+\gamma_3.
\end{equation}
For a pure Bloch vector ${\bm r}$, the escape rate is $\sum_j\gamma_j(1-x_j^2)$. Since \(\sum_j x_j^2=1\),
\[
\sum_{j=1}^3\gamma_j(1-x_j^2)
=\gamma_1+\gamma_2+\gamma_3-\sum_{j=1}^3\gamma_jx_j^2.
\]
The ordering implies $\sum_{j=1}^3\gamma_jx_j^2\ge\gamma_3$, with equality at \({\bm r}=(0,0,\pm1)\). Therefore
\[
\nu_{q,3} = \max_{\|{\bm r}\|=1}
\sum_{j=1}^3\gamma_j(1-x_j^2)=\gamma_1+\gamma_2 ,
\]
and hence $\nu_{q,3}=R$. The Pauli matrices have Hilbert--Schmidt norm $\sqrt2$, so the canonical rates are $2\gamma_j$ and $\nu_{q,1}=2\sum_j\gamma_j$; this factor must be kept in mind when the rates $\gamma_j$ of \eqref{eq:pauligen} are compared with $\nu_{q,1}$.

\begin{proposition}\label{prop:pauli}
For the Pauli generator above,
\begin{equation}
{\nu_{\mathrm{CP}}(\mathcal L)
=\nu_{q,2} 
= \gamma := \sum_{i=1}^3\gamma_i.}
\end{equation}
The infimum is attained at $G=\gamma I$.
\end{proposition}

\begin{proof}
The canonical jump map and the corresponding operator $G$ are
\begin{equation*}
\Phi_{\mathrm{can}}(X)=\sum_i\gamma_i\sigma_iX\sigma_i,
\qquad
G_{\mathrm{can}}=\Phi_{\mathrm{can}}^\dagger(I)=\gamma I.
\end{equation*}
Consequently, this is a feasible representation and
\begin{equation}
\nu_{\mathrm{CP}}(\mathcal L)
\leq\|G_{\mathrm{can}}\|_\infty
=\gamma . %=\nu_{2,\mathrm{can}}(\mathcal L).
\end{equation}

To prove the reverse inequality, let $G$ be any feasible Hermitian
operator. We use the unnormalized maximally entangled vector and
Choi operator
\begin{equation*}
|\Gamma\rangle=|00\rangle+|11\rangle,\qquad
\langle\Gamma|\Gamma\rangle=2,\qquad
C(\Psi)=(\id_2\otimes\Psi)(|\Gamma\rangle\langle\Gamma|).
\end{equation*}
For any $2\times2$ matrix $B$,
\begin{equation*}
\langle\Gamma|(I\otimes B)|\Gamma\rangle=\Tr B.
\end{equation*}
Put $|\Gamma_i\rangle=(I\otimes\sigma_i)|\Gamma\rangle$.
Since $\Tr\sigma_i=0$, we have
$\langle\Gamma|\Gamma_i\rangle=0$, and hence
\begin{align}
C(\mathcal L)
=\sum_i\gamma_i|\Gamma_i\rangle\langle\Gamma_i|
-\gamma|\Gamma\rangle\langle\Gamma|,\quad 
\langle\Gamma|C(\mathcal L)|\Gamma\rangle =-4\gamma.
\end{align}
The Choi operator of the anticommutator map is
\begin{equation}
C\left(\{G,\cdot\}\right)
=
(I\otimes G)|\Gamma\rangle\langle\Gamma|
+|\Gamma\rangle\langle\Gamma|(I\otimes G) .
\end{equation}
Its expectation in the same unnormalized vector equals
\begin{equation}
\left\langle\Gamma\left|
C\left(\{G,\cdot\}\right)
\right|\Gamma\right\rangle
=4\Tr G.
\end{equation}
Complete positivity implies $C(\Phi_G)\geq0$. Therefore
\begin{equation}
0\leq\langle\Gamma|C(\Phi_G)|\Gamma\rangle
=-4\gamma+2\Tr G,
\end{equation}
and hence $ \Tr G\geq2\gamma$. Because $G\geq0$ acts on a two-dimensional space,
\begin{equation}
\|G\|_\infty\geq\frac{\Tr G}{2}\geq \gamma.
\end{equation}
This holds for every feasible $G$ and proves the claim.
\end{proof}
As a result we obtain
\begin{equation}\label{eq:pauli}
R=\nu_{q,3}=\gamma_1+\gamma_2\;\le\;\nu_\CP=\nu_{q,2}=\gamma_1+\gamma_2+\gamma_3,
\end{equation}
with strict inequality whenever $\gamma_3>0$.

\end{example}

\begin{example}[Depolarization]\label{ex:depol}
For
\begin{equation}\label{eq:depol}
\cL_{\rm dep}(X)=\gamma\Bigl(\tr(X)\frac{I}{n}-X\Bigr),\qquad\gamma>0,
\end{equation}
the only nonzero eigenvalue is $-\gamma$ (on traceless $X$), so $R=\gamma/2$, while $\nu_{q,3}=\gamma(1-1/n)$. With an orthonormal basis $\{W_\alpha\}_{\alpha=0}^{n^2-1}$ of $M_n$, $W_0=I/\sqrt n$, one has $\tr(X)I/n=\frac1n\sum_\alpha W_\alpha XW_\alpha^\dagger$, so the canonical jumps are $\sqrt{\gamma/n}\,W_\alpha$, $\alpha\ge1$, and $G_{\rm can}=\gamma(1-1/n^2)I$. Conversely, $C(\cL_{\rm dep})=\gamma(I_{n^2}/n-P_\Gamma)$, and evaluating CP feasibility on $\Gamma$ gives $0\le\gamma(1-n^2)+n\tr G$, hence $\|G\|_\infty\ge\tr G/n\ge\gamma(1-1/n^2)$. Thus
\begin{equation}\label{eq:depolcp}
R=\frac\gamma2,\qquad \nu_{q,3}=\gamma\Bigl(1-\frac1n\Bigr),\qquad \nu_\CP=\nu_{q,2}=\gamma\Bigl(1-\frac1{n^2}\Bigr).
\end{equation}
For $n>2$ the escape-rate disk is valid but not optimal, $R<\nu_{q,3}$; already for a qubit the optimized CP scale exceeds the exact spectral scale, $\nu_\CP=3\gamma/4>R=\gamma/2$.
\end{example}

\begin{example}[Recovery of the classical exit-rate parameter]\label{sec:embedding}

For a column Kolmogorov generator $K$, use jump
$V_{ij}=\sqrt{K_{ij}}\ket i\bra j$ for $i\ne j$. One has

\begin{equation}\label{LK}
    \cL_K(\rho) = \sum_{i\neq j} K_{ij} E_{ij} \rho E_{ji} - \frac 12 \{G,\rho\} ,
\end{equation}
with
\begin{equation}
G =\sum_j(-K_{jj})\ket j\bra j.
\end{equation}
On diagonal density matrices the resulting GKLS generator $\cL_K$ acts as $K$. 
The jumps $\ket i\bra j$, $i\ne j$, are traceless and Hilbert--Schmidt orthonormal, so this is the canonical representation and $\nu_{q,2}=\max_j(-K_{jj})=\nu_c$ is intrinsic to $\cL_K$.

\begin{proposition}[Classical embedding]\label{prop:embed}
For a GKLS generator (\ref{LK}) one has 

$$ \nu_{q,3}(\cL_K)=\nu_\CP(\cL_K)=\nu_{q,2}=\nu_c , $$
where $\nu_c = \max_i |K_{ii}|$. 
\end{proposition}
\begin{proof}
Basis states give $\nu_{q,3}\ge|\bra j\cL_K(\ket j\bra j)\ket j|=|K_{jj}|$, hence $\nu_{q,3}\ge\nu_c$. Proposition~\ref{prop:lower}, applied to every feasible representation, and Proposition~\ref{prop:attainment} give $\nu_{q,3}\le\nu_\CP\le\nu_{q,2}=\nu_c$.
\end{proof}
The quantum rate bound is thus an exact extension of the classical exit-rate parameter. It bounds the population spectrum and the additional coherence modes, with eigenvalues $-(w_i+w_j)/2$, $w_j=-K_{jj}$, simultaneously, and for this embedding the escape rate and the representation scales $\nu_\CP,\nu_{q,2}$ collapse to the classical value $\nu_c$ (whereas $R(\cL_K)\le\nu_c$ and $\nu_{q,1}=\sum_j|K_{jj}|\ge\nu_c$ in general).

\end{example}

\begin{table}[t]
\centering\small
\setlength{\tabcolsep}{4pt}\begin{tabular}{p{.36\linewidth}cccc}
\toprule
Generator & $R$ & $\nu_{q,3}$ & $\nu_{\CP}$ & $\nu_{q,2}$\\
\midrule
Hermitian single jump, spectral diameter $\Delta=a_+-a_-$ & $\Delta^2/4$ & $\Delta^2/4$ & $\Delta^2/4$ & $\max\{a_+^2,a_-^2\}$\\[2mm]
Qutrit counterexample \eqref{eq:A} & $10$ & $9$ & $10$ & $10$\\[2mm]
Qutrit counterexample \eqref{eq:best} & $8$ & $64/9$ & $8$ & $8$\\[2mm]
Pauli \eqref{eq:pauligen} & $\gamma_1+\gamma_2$ & $\gamma_1+\gamma_2$ & $\sum_j\gamma_j$ & $\sum_j\gamma_j$\\[2mm]
Depolarization \eqref{eq:depol} & $\gamma/2$ & $\gamma(1-\frac1n)$ & $\gamma(1-\frac1{n^2})$ & $\gamma(1-\frac1{n^2})$\\[2mm]
Classical embedding $\cL_K$ & $\le\nu_c$ & $\nu_c$ & $\nu_c$ & $\nu_c$\\
\bottomrule
\end{tabular}
\caption{Exact spectral scale, escape rate, optimized CP representation scale and canonical scale for the examples of Sections~\ref{sec:false} and~\ref{sec:optimization}. For the Hermitian jump, $a_k$ are the eigenvalues of the traceless jump. In all cases $\max\{R,\nu_{q,3}\}\le\nu_\CP\le\nu_{q,2}$; there is no universal ordering between $R$ and $\nu_{q,3}$.}\label{tab:scales}
\end{table}

\section{Beyond the disk: Brauer--Cassini ovals and thermodynamic ellipses}\label{sec:cassini-thermo}

It turns out that the spectral disk $D(-\nu_c,\nu_c)$ implied by the Gershgorin theorem for any Kolmogorov generator $K$ can be substantially improved. The classical tangent disk can be refined in two different ways. Brauer's localization uses
two escape rates, while thermodynamic ellipses use stationary fluxes or cycle affinities.
These refinements are complementary and clarify which additional structure a quantum analogue
would have to capture.

\subsection{Brauer--Cassini ovals}\label{sec:cassini}

\begin{theorem}[Brauer's Cassini oval theorem \cite{Brauer1946, Brauer1947, Varga2004}]
Let $A=(a_{ij})\in M_n(\mathbb C)$, with $n\geq2$, and define
the off-diagonal row sums
\[
R_i=\sum_{k\neq i}|a_{ik}|.
\]
Then
\[
\spec(A)\subseteq\bigcup_{1\leq i<j\leq n}\mathcal C_{ij}(A),
\]
where
\[
\mathcal C_{ij}(A)
=
\left\{
z\in\mathbb C:
|z-a_{ii}|\,|z-a_{jj}|\leq R_iR_j
\right\}.
\]
These sets are the filled Cassini ovals with foci $a_{ii}$
and $a_{jj}$.
\end{theorem}

Brauer's theorem strengthens the Gershgorin inclusion. Indeed, with the disks $D(a,r)$ of the Introduction,
\[
\mathcal C_{ij}(A)
\subseteq D(a_{ii},R_i)\cup D(a_{jj},R_j).
\]
To see this, a point outside both disks satisfies
$|z-a_{ii}|>R_i$ and $|z-a_{jj}|>R_j$, and hence violates
the defining inequality of $\mathcal C_{ij}(A)$.
Consequently,
\[
\spec(A)
\subseteq\bigcup_{i<j}\mathcal C_{ij}(A)
\subseteq\bigcup_iD(a_{ii},R_i).
\]

\begin{theorem}[A single Cassini oval for stochastic matrices \cite{Varga2004}]
Let $P=(p_{ij})\in M_n(\mathbb R)$, $n\geq2$, be a column-
or row-stochastic matrix. Relabel the indices so that
\[
p_{11}\leq p_{22}\leq\cdots\leq p_{nn},
\]
and define its Brauer--Cassini regions by
\[
\mathcal C_{ij}(P)
=
\left\{
z\in\mathbb C:
|z-p_{ii}|\,|z-p_{jj}|
\leq(1-p_{ii})(1-p_{jj})
\right\},
\qquad i<j.
\]
Then
%\[
%\bigcup_{i<j}\mathcal C_{ij}(P)=\mathcal C_{12}(P).
%\]
%In particular,
\[
{
\spec(P) \, \subseteq\,  \mathcal C_{12}(P) 
%= \left\{
%z\in\mathbb C:
%|z-p_{11}|\,|z-p_{22}|
%\leq(1-p_{11})(1-p_{22}) 
%\right\} 
\, \subseteq \, D(0,1) .
}
\]
Thus the Cassini region determined by the two smallest diagonal
entries contains all the other Brauer--Cassini regions.
\end{theorem}
The theorem follows from Brauer's theorem, since for ordered diagonal entries the region $\mathcal C_{12}(P)$ contains all other $\mathcal C_{ij}(P)$; see \cite{Varga2004}. The ordering is essential: without it the regions are not nested.

\begin{corollary}[A single Cassini oval for Markov generators]
Let $K\in M_n(\mathbb R)$, $n\geq2$, be a continuous-time
Markov generator in the column convention:
\[
K_{ij}\geq0\quad(i\neq j),
\qquad
\sum_iK_{ij}=0.
\]
Order its escape rates as
\[
\nu_c := r_1\geq r_2\geq\cdots\geq r_n\geq0,
\qquad r_i=-K_{ii}.
\]
Then
\begin{equation}  \label{eq:cassini}
{
\spec(K)\subseteq
\mathcal C(r_1,r_2):=\left\{
z\in\mathbb C:
|z+r_1|\,|z+r_2|\leq r_1r_2
\right\} \subseteq D(-r_1,r_1) . 
}
\end{equation}
\end{corollary}
\begin{proof}
For $K=0$ the statement is trivial. Otherwise $P=I+K/r_1$ is column-stochastic with $p_{ii}=1-r_i/r_1$, ordered as in the theorem, and $\lambda=r_1(\mu-1)$ maps $\mathcal C_{12}(P)$ onto $\mathcal C(r_1,r_2)$.
\end{proof}

\subsection{Thermodynamic ellipses}\label{sec:ellipses}

%\paragraph{The edge-force ellipse.}
Assume now that $K$ is irreducible with stationary distribution $\pi>0$ and
bidirectional transition support. Define
\[
F_{ij}=\log\frac{K_{ij}\pi_j}{K_{ji}\pi_i},
\qquad
\beta_{\rm e}=\tanh\!\left(\frac12\max_{K_{ij}>0}|F_{ij}|\right),
\qquad
\Pi=\operatorname{diag}(\pi).
\]
Xu et al.~\cite{Xu2025} prove
\begin{equation}\label{eq:edge-ellipse}
\spec K\subseteq W(\Pi^{-1/2}K\Pi^{1/2})\subseteq
\mathcal E(r_1,\beta_{\rm e}),
\end{equation}
where
\begin{equation}
\mathcal E(r,\beta):=
\left\{x+iy:\frac{(x+r)^2}{r^2}+\frac{y^2}{r^2\beta^2}\le1\right\},
\end{equation}
and $W(A)$ denotes the numerical range of $A$, that is, $W(A)=\{\langle x,Ax\rangle:\|x\|=1\}$ \cite{HornJohnson2013}. 
At $\beta=0$ we interpret the ellipse as $[-2r,0]$. If some transition is 
unidirectional (meaning $K_{ij} > 0$ but $K_{ji}=0$), which is excluded by the
bidirectional assumption above, one may set $|F_{ij}|=\infty$ and $\beta_{\rm e}=1$, so that the ellipse becomes
the disk $D(-r_1,r_1)$ and the statement reduces to Gershgorin disk. Classical detailed balance ($F_{ij}=0$ for all edges)
collapses the ellipse to the real segment $[-2r_1,0]$.
For $\lambda=-\Gamma+i\omega$, the restriction is
$\omega^2\le\beta_{\rm e}^2\Gamma(2r_1-\Gamma)$.
Combining \eqref{eq:cassini} and \eqref{eq:edge-ellipse} gives the following bound for the spectrum of $K$

\begin{equation}
    \spec K\subseteq\mathcal C(r_1,r_2)\cap\mathcal E(r_1,\beta_{\rm e}) .
\end{equation}
The two regions reflect different features of the dynamics: the Brauer-Cassini oval depends on the two largest exit rates, \(r_1\) and \(r_2\), whereas the thermodynamic ellipse depends on the imbalance between forward and backward stationary flows, measured by the edge forces \(F_{ij}\).

%\paragraph{The Uhl--Seifert ellipse.}
Under the same assumptions (irreducibility and bidirectional transitions), for an oriented simple cycle $c$ of length $n_c\ge3$, define
\[
A_c=\sum_{(j\to i)\in c}\log\frac{K_{ij}}{K_{ji}},
\qquad
\beta_{\rm c}=\max_c\tanh\frac{|A_c|}{2n_c},
\]
with $\beta_{\rm c}=0$ if no such cycles exist.
Uhl and Seifert~\cite{UhlSeifert2019} conjectured
$\spec K\subseteq\mathcal E(r_1,\beta_{\rm c})$.
As an eigenvalue enclosure this is sharper, since $A_c=\sum_cF_{ij}$ gives $\beta_{\rm c}\le\beta_{\rm e}$;
the result of \cite{Xu2025} also covers the numerical range. Although open in \cite{Xu2025}, the conjecture is
stated to be proved in the later preprint \cite{Kolchinsky2026} (Theorem~3 and equation~(8) of arXiv v3);
we have not independently verified that proof.

\begin{figure}[ht]
\centering
\includegraphics[width=0.7\textwidth]{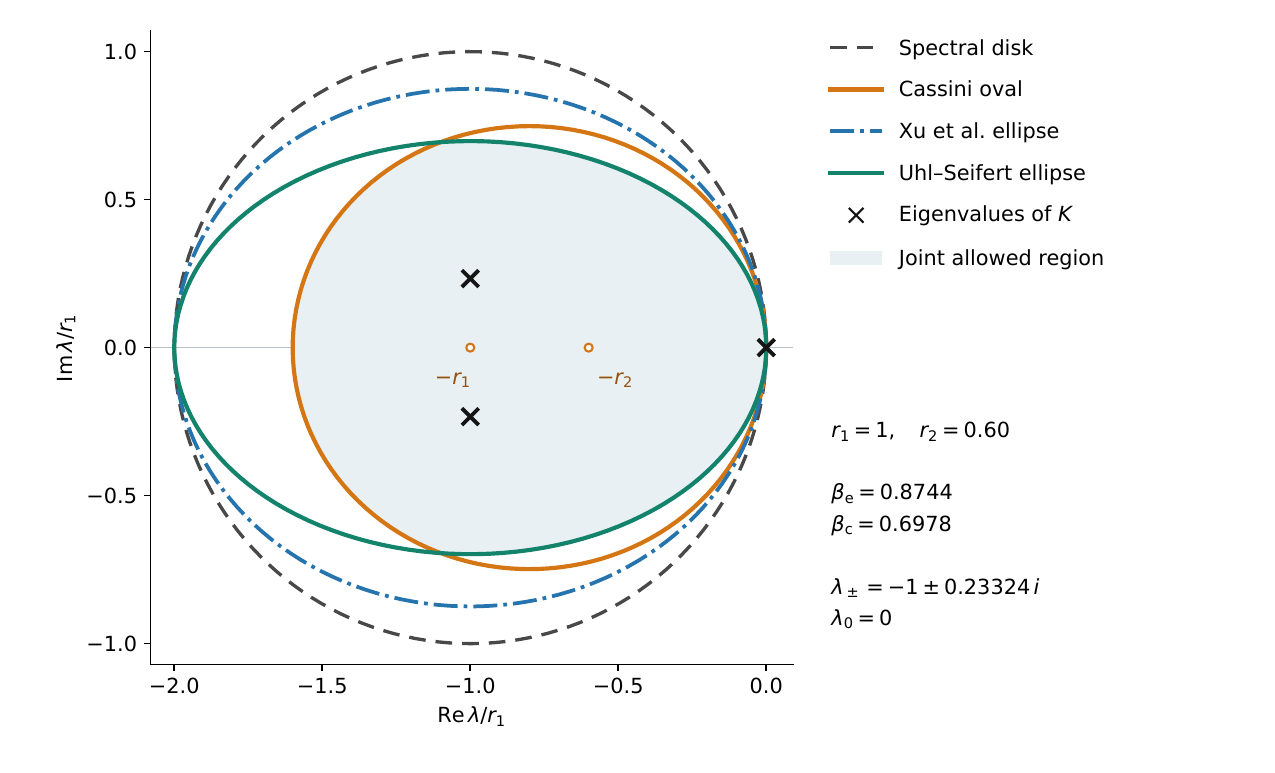}
\caption{Spectral enclosures for the three-state generator (\ref{K3}). 
The spectral disk $D(-r_1,r_1)$ (gray dashed), Brauer's Cassini oval
$|z+r_1|\,|z+r_2|\le r_1r_2$ (orange), the edge-force ellipse of
Xu et al.~\cite{Xu2025} (blue dash-dotted), and the Uhl--Seifert
cycle-affinity ellipse~\cite{UhlSeifert2019,Kolchinsky2026} (green)
are shown for the same generator.
Here $r_1=1$, $r_2=0.60$, $\beta_{\rm e}\simeq0.87438$, and
$\beta_{\rm c}\simeq0.69780$.
Both ellipses have center $-r_1$ and horizontal semiaxis $r_1$;
their vertical semiaxes are $r_1\beta_{\rm e}$ and
$r_1\beta_{\rm c}$.
Crosses mark the exact eigenvalues $0$ and $-1\pm i\sqrt{34}/25$.
The shaded region is the intersection of all four enclosures,
equivalently the intersection of the Cassini region and the Uhl--Seifert ellipse.
The small open circles indicate the Cassini foci.
The disk is the classical specialization of our spectral-disk bound;
the thermodynamic refinements apply here to the classical rate matrix.}
\label{fig:spectral-enclosures}
\end{figure}

To exhibit the final proof step, we quote (without re-deriving it) the cycle-localization theorem of
\cite{Kolchinsky2026} in its special case with interpolation parameter $1/2$, in the notation of that preprint:
under the assumptions above it supplies, for each nonreal $\lambda=-\Gamma+i\omega$, a cycle $c$
and $\theta\in(0,\pi/2)$ such that
\[
\frac12\left(\frac{|\omega|}{\Gamma}\tan\theta+
\frac{|\omega|}{2r_1-\Gamma}\cot\theta\right)
\le\tanh\frac{|A_c|}{2n_c}.
\]
By Gershgorin, $0<\Gamma<2r_1$ for nonreal $\lambda$, so both coefficients are positive, and the
arithmetic--geometric mean inequality bounds the left side below by
$|\omega|/\sqrt{\Gamma(2r_1-\Gamma)}$, uniformly in $\theta$. Hence $\omega^2\le\beta_{\rm c}^2\Gamma(2r_1-\Gamma)$,
which is the conjectured ellipse.
Real eigenvalues are covered by Gershgorin. This derivation uses their
cycle-localization theorem; it is not an independent proof of that theorem.
Figure~\ref{fig:spectral-enclosures} compares these enclosures for a single three-state generator.

\begin{equation}  \label{K3}
    K=\begin{pmatrix}
-1&0.12&0.28\\
0.95&-0.60&0.12\\
0.05&0.48&-0.40
\end{pmatrix} .
\end{equation}

%\paragraph{Implications for the quantum problem.}

These classical results motivate a refinement involving both kinetic and reversibility data,
rather than a single scalar rate. They do not directly yield a quantum ellipse:
a classical compression $K_{ij}=\tr[P_i\cL(P_j)]$ need not contain the complex
spectrum of $\cL$. For comparison, a real eigenvalue $\lambda$ admits a nonzero Hermitian
eigenoperator $X=\sum_jx_jP_j$. The compressed matrix
$K^{(X)}_{ij}:=\tr[P_i\cL(P_j)]$ is a Kolmogorov generator and satisfies
\[
(K^{(X)}x)_i=\tr[P_i\cL(X)]=\lambda x_i.
\]
This is the superdecoherence of $\cL$ in the eigenbasis of $X$ \cite{superdecoherence}.
The reduction is therefore available for each real eigenvalue, but not in general for nonreal
modes. The diagonal-jump counterexample
already rules out a universal ellipse contained in $D(-\nu_{q,3},\nu_{q,3})$: the eigenvalues $-18\pm6i$ have real part $-2\nu_{q,3}$ and nonzero imaginary part, so they lie outside every ellipse $\mathcal E(\nu_{q,3},\beta)$.
Even the Cassini refinement does not transfer to the quantum embedding $\cL_K$ of
Section~\ref{sec:embedding}, although there $\nu_{q,3}=\nu_c=r_1$: its coherence eigenvalue
$-(r_1+r_2)/2$ lies in $\mathcal C(r_1,r_2)$ if and only if $r_1/r_2\le3+2\sqrt2$.

\section{Conclusions and open questions}\label{sec:conclusions}

The spectral disk problem has a direct solution. For
$\mathcal L=\Phi-\tfrac12\{\Phi^\dagger(I),\cdot\}$, complete positivity of $\Phi$, and in fact already its positivity, places the spectrum in the
tangent disk with scale $\|\Phi^\dagger(I)\|_\infty$. Maximal singular vectors provide the quantum counterpart
of the maximal-component argument for classical generators. This proof establishes the first two canonical
scaling conjectures of \cite{corralling}.

The third conjecture separates two effects that coincide classically. Pure-state escape controls initial
loss of survival probability, but a spectral disk also constrains oscillation frequencies.
The qutrit counterexample has vanishing canonical Hamiltonian, a normal jump and a normal unital generator;
its violation cannot be attributed to a noncanonical coherent term or to nonnormal transient behavior.
The escape rate nevertheless controls every decay rate of an arbitrary positive trace-preserving
semigroup: $-\Re\lambda\le2\nu_{q,3}$ by Theorem~\ref{thm:universal-decay-rate}.
The coefficient two is optimal, and no unitality, complete-positivity, or zero-Hamiltonian
assumption is needed. This is precisely the real-part restriction imposed by the escape-rate
disk, but it does not imply the additional bound on the imaginary part
(Remark~\ref{rem:decay-vs-disk}). Consequently, every real eigenvalue lies in
$[-2\nu_{q,3},0]$, and the full escape-rate disk contains the spectrum whenever it is real.
Besides Hilbert--Schmidt self-adjoint generators and qubit generators with vanishing canonical
Hamiltonian, this includes generators whose entire Heisenberg generator is self-adjoint in a
weighted inner product associated with a faithful state. 
The disk inclusion need not be optimal: for depolarization in dimension greater than two,
$R<\nu_{q,3}$.

For GKLS generators with vanishing canonical Hamiltonian, Theorem~\ref{thm:sharp4} gives the
complementary estimate $R\le\nu_{q,2}\le4\nu_{q,3}$. Thus an escape-rate disk enlarged by a
factor of four always contains the full spectrum. The traceless single-jump family
\eqref{eq:Anb} shows that this factor cannot be reduced uniformly over all finite dimensions.
This resolves the dimension-independent disk question without asserting that four is optimal
at each fixed dimension.

Representation optimization addresses a separate question. For every GKLS generator with vanishing canonical
Hamiltonian the proved hierarchy is
\[
\max\{R(\mathcal L),\nu_{q,3}\}
\le\nu_{\CP}\le\nu_{q,2}\le\nu_{q,1}.
\]
There is no universal ordering between $R$ and $\nu_{q,3}$: suitable diagonal qutrit jumps, such as \eqref{eq:A}, give $R>\nu_{q,3}$,
whereas depolarization in dimension greater than two gives $R<\nu_{q,3}$.
Moreover, depolarization shows that the optimized CP representation can lie strictly
above both quantities. Hence canonical normalization, optimal jump-map representation, and optimal spectral
normalization should not be identified.

 For the canonical quantum embedding
of a Kolmogorov generator, the escape rate and the representation scales $\nu_\CP$ and $\nu_{q,2}$
reduce to the largest classical exit rate, while the total rate $\nu_{q,1}$ is in general larger.
The equality is special to this structure. In general quantum dynamics, coherence modes and the freedom
to choose the jump map produce gaps that have no counterpart in the classical case.
The family of qutrit counterexamples in Appendix~\ref{sec:geometry} makes this explicit.

Classical Cassini ovals and thermodynamic ellipses (Section~\ref{sec:cassini-thermo}) show that
sharper enclosures require more information than a single rate. The Cassini oval uses the two largest exit
rates; the edge-force ellipse of \cite{Xu2025} bounds oscillation frequencies by the largest thermodynamic
force; the Uhl--Seifert ellipse \cite{UhlSeifert2019,Kolchinsky2026} replaces it by cycle affinities and
collapses to the real axis at detailed balance. On the quantum side three obstructions appear. A classical
compression can be adapted to a real eigenoperator but does not generally capture nonreal
quantum modes (Section~\ref{sec:ellipses}); the Cassini oval already
fails for the quantum embedding of a Kolmogorov generator once $r_1/r_2>3+2\sqrt2$; and the qutrit
counterexample rules out every ellipse centred at $-\nu_{q,3}$ with horizontal semiaxis $\nu_{q,3}$. A quantum thermodynamic
enclosure must therefore combine a representation scale, such as $\|G\|_\infty$, with an
intrinsic measure of deviation from detailed balance. Our detailed-balance and real-spectrum results give
its reversible endpoint, where the enclosure reduces to the segment $[-2\nu_{q,3},0]$.

Several questions follow naturally from these results. It would be interesting to classify generators
for which $R=\nu_{\CP}$ holds and to characterize equality in the universal decay-rate bound.
While Theorem~\ref{thm:sharp4} determines the optimal dimension-independent coefficient,
the optimal coefficient in $R\le c_n\nu_{q,3}$ at each fixed dimension remains to be determined
beyond the established cases. Finally, it would be interesting to analyze a quantum analogue
of the edge-force ellipse, with horizontal scale given by a representation scale such as
$\|G\|_\infty$ and vertical scale controlled by ratios of stationary jump fluxes.

\section*{Acknowledgments}

We are grateful to Wenlin Zhang for a message informing
 us about the counterexample to Conjecture 3 which triggered this project.
D.C. was supported by the Polish National Science Center, project
2024/55/B/ST2/01781 while KZ acknowledges funding
by the European Union under ERC Advanced Grant TAtypic, Project No. 101142236. This work  also benefited from discussions and meetings  (D. C. and S. D.) within COST Action CA24109, Many-body Open Quantum Systems (QOpen), activities, supported by COST: European Cooperation in Science and Technology.

\appendix

\numberwithin{equation}{section}

\section{Spectral disk for positive jump maps}\label{app:lemma}

Here we show that complete positivity in Theorem~\ref{thm:disk} can be relaxed to positivity. The Cauchy--Schwarz step \eqref{eq:CS} is replaced by the following matrix-element bound.

\begin{theorem}[Spectral disk for positive jump maps]\label{thm:diskpos}
Let $\Phi:M_n\to M_n$ be a positive linear map, $G=\Phi^\dagger(I)$, and $\cL(X)=\Phi(X)-\frac12(GX+XG)$. Then $\spec(\cL)\subseteq D(-\|G\|_\infty,\|G\|_\infty)$.
\end{theorem}
\begin{proof}
The adjoint of a positive map is positive: for $X,Y\ge0$, $\tr[X\Phi^\dagger(Y)]=\tr[\Phi(X)Y]\ge0$; in particular $G=\Phi^\dagger(I)\ge0$. The proof of Theorem~\ref{thm:disk} then applies verbatim, except that \eqref{eq:CS} follows from Lemma~\ref{lem:positive} stated below
and applied to $\Psi=\Phi^\dagger$, $\Psi(I)=G$.
\end{proof}
%% now the narration is straight and simpler to follow (KZ)

\begin{lemma}\label{lem:positive}
Let $\Psi:M_n\to M_n$ be positive and put $G=\Psi(I)$. For all $X\in M_n$ and $u,v\in\mathbb C^n$,
\begin{equation}\label{eq:element}
|\langle u,\Psi(X)v\rangle|
\le\|X\|_\infty
\sqrt{\langle u,Gu\rangle\langle v,Gv\rangle}.
\end{equation}
\end{lemma}

\noindent
\emph{Proof}. 
%of Lemma~\ref{lem:positive}.} 
The proof reduces the bound to contractivity of positive unital maps.
Let $T(I)=I$ and $T$ be positive. For a unitary $U=\sum_j z_jP_j$, with $|z_j|=1$,
put $E_j=T(P_j)$. These operators are positive and sum to $I$.
The map $Wx=(E_j^{1/2}x)_j$ is therefore an isometry, since
$\|Wx\|^2=\sum_j\langle x,E_jx\rangle=\|x\|^2$. Moreover,
\[
T(U)=W^*\operatorname{diag}(z_jI)W,
\qquad \|T(U)\|_\infty\le1.
\]
For a general contraction $X$, take a polar decomposition $X=UH$ with $0\le H\le I$.
Because the matrices are square and finite dimensional, the partial isometry can be extended to a unitary $U$.
The matrix $Z=H+i\sqrt{I-H^2}$ is unitary, and $H=(Z+Z^*)/2$.
Thus $X=(UZ+UZ^*)/2$ is an average of two unitaries and $\|T(X)\|_\infty\le1$.
Scaling proves contractivity for arbitrary $X$. This also follows from the Russo--Dye theorem \cite{russodye}.

If $G=\Psi(I)>0$, then $T(X)=G^{-1/2}\Psi(X)G^{-1/2}$ is positive and unital.
Consequently
\begin{align*}
|\langle u,\Psi(X)v\rangle|
=|\langle G^{1/2}u,T(X)G^{1/2}v\rangle| \le\|X\|_\infty\|G^{1/2}u\|\|G^{1/2}v\|,
\end{align*}
which is the desired bound.
%\emph{Step 3: remove invertibility.}
For $\varepsilon>0$ use the positive map
$\Psi_\varepsilon(X)=\Psi(X)+\varepsilon\tr(X)I/n$.
Its value at $I$ is $G+\varepsilon I>0$, so Step 2 applies.
Both sides of the resulting scalar inequality converge as $\varepsilon\downarrow0$.
This proves the claim even if one of the two expectations on the right vanishes, without inserting an inverse of a singular operator.

\section{A family of qutrit counterexamples}\label{sec:geometry}
\begin{figure}[t]
\centering
\includegraphics[width=0.5\textwidth]{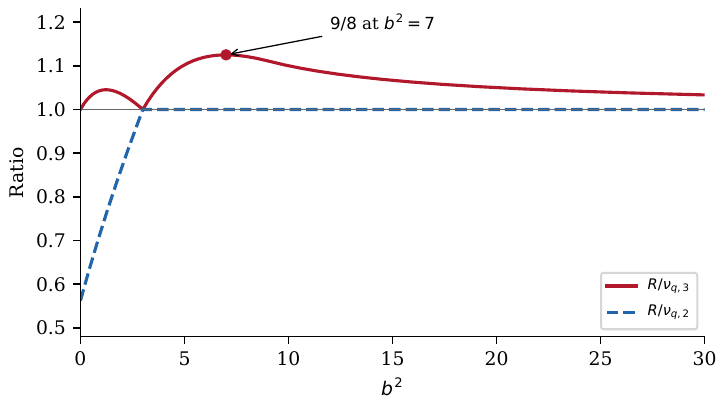}
\caption{$R/\nu_{q,3}$ and $R/\nu_{q,2}$ along $A_b=\operatorname{diag}(1+ib,1-ib,-2)$, equation~\eqref{eq:familyratio}. Conjecture 3 fails for every $b\ne0$ except $b^2=3$, maximally at $b^2=7$. Example~\ref{ex:counter} is the case $b=3$.}\label{fig:family}
\end{figure}

Consider the traceless normal jump
\[
A_b=\operatorname{diag}(1+ib,\,1-ib,\,-2),\qquad b\in\mathbb R,\qquad s=b^2,
\]
and $\cL=\D_{A_b}$, which is in canonical form with vanishing Hamiltonian; Example~\ref{ex:counter} is the case $b=3$. By \eqref{eq:diagonaleigen}, the nonzero eigenvalues are
\[
\lambda_{12}=-2s+2ib,\qquad \lambda_{13}=-\frac{9+s}{2}-2ib,\qquad \lambda_{23}=\overline{\lambda_{13}},
\]
together with their complex conjugates (on $E_{21},E_{31},E_{32}$). Their contributions $|\lambda|^2/(-2\Re\lambda)$ are $1+s$ and $\frac{9+s}{4}+\frac{4s}{9+s}$, hence
\[
R=\max\Bigl\{1+s,\ \tfrac{9+s}{4}+\tfrac{4s}{9+s}\Bigr\},
\]
and the two candidates exchange dominance at $s=3$, since their difference is $3(s-3)(s+5)/[4(9+s)]$.
 Since $A_b$ is diagonal, by \eqref{eq:variance} the escape rate depends only on $p=|\psi_1|^2$, $q=|\psi_2|^2$, $r=|\psi_3|^2$ and equals the variance of the points $1\pm ib,-2$:
\[
s+(9-s)r-9r^2-s(p-q)^2 .
\]
At fixed $r$ it is maximized by $p=q=(1-r)/2$, and the maximizing $r$ is $(9-s)/18$ for $s\le9$ and $0$ for $s\ge9$. Thus
\[
\nu_{q,3}=\begin{cases}(9+s)^2/36,&s\le9,\\ s,&s\ge9,\end{cases}\qquad
\nu_{q,2}=\max\{1+s,4\},\qquad \nu_{q,1}=6+2s.
\]
%\emph{Violation of Conjecture 3.} 
Consequently
\begin{equation}\label{eq:familyratio}
\frac{R}{\nu_{q,3}}=\begin{cases}\dfrac{9\,[(9+s)^2+16s]}{(9+s)^3},&0<s\le3,\\[8pt] \dfrac{36(1+s)}{(9+s)^2},&3\le s\le9,\\[8pt] 1+\dfrac1s,&s\ge9.\end{cases}
\end{equation}
The numerator of $R/\nu_{q,3}-1$ is $s(3-s)(s+21)$ on the first branch and $(s-3)(15-s)$ on the second, and the difference is $1/s$ on the third. Thus Conjecture~3 fails for every $b\ne0$ except $b^2=3$. The two exceptions are the Hermitian jump ($b=0$, real spectrum) and the equilateral configuration ($b^2=3$), where the points $a_k$ lie on a circle centered at their mean and $R=\nu_{q,3}=\nu_{q,2}=4$. For $s\ge3$ one has $R=\nu_{q,2}$, so Theorem~\ref{thm:disk} is saturated along the whole branch. The second branch is maximal at $s=7$, and the other two branches %stay below  (KZ)
do not exceed $10/9$, so the largest violation in the family is

\[ A=\operatorname{diag}(1+i\sqrt7,\,1-i\sqrt7,\,-2),\]
giving rise to 
\begin{equation}\label{eq:best}
\nu_{q,3}=\tfrac{64}{9},\qquad R=\nu_{q,2}=\nu_\CP=8,\qquad
\nu_{q,1}=20,\qquad \frac{R}{\nu_{q,3}}=\frac98 ,
\end{equation}
where $\nu_\CP=8$ follows from \eqref{eq:cpchain}. Its nonzero eigenvalues are $-14\pm2i\sqrt7$ (simple) and $-8\pm2i\sqrt7$ (each of multiplicity two). Figure~\ref{fig:family} shows the whole family.

\section{Proof of Lemma \ref{lem:four}} \label{App-4nu}

{

The case $n=1$ is trivial. Fix a unit vector $e_i$ and extend it to an
orthonormal basis $\{e_j\}_{j=1}^n$. Define vectors $u_j \in \mathbb{C}^m$ and nonnegative numbers $q_j$
\[
(u_j)_\alpha :=\langle e_j,V_\alpha e_j\rangle,\quad \alpha=1,\ldots,m,
\qquad q_j:=\sum_\alpha\sum_{k\ne j}|(V_\alpha)_{kj}|^2.
\]
For a unit vector $\psi\in\mathbb C^n$, write
\[
e(\psi):=\sum_\alpha\bigl(\|V_\alpha\psi\|^2
-|\langle\psi,V_\alpha\psi\rangle|^2\bigr).
\]
This is the escape rate for the generator with these jumps. In particular,
$q_j=e(e_j)$ and
$\langle e_j,G e_j\rangle=\|u_j\|^2+q_j$. 
Tracelessness gives $\sum_j u_j=0$, so there exists $j\ne i$ with
$\Re \langle u_i,u_j\rangle\leq0$. Let
\[
\psi_\theta=\frac{e_i+e^{i\theta}e_j}{\sqrt2},\qquad
\theta\in\{0,\pi/2,\pi,3\pi/2\},
\]
and consider

\[ \langle e(\psi_\theta)\rangle := \frac 14 \big(  e(\psi_{\theta=0}) +  e(\psi_{\theta=\pi/2}) +  e(\psi_{\theta=\pi}) + e(\psi_{\theta=3\pi/2}) \big) .   \]
 Direct expansion yields
\begin{align}
\langle e(\psi_\theta)\rangle 
 &=\tfrac14\|u_i-u_j\|^2+\tfrac12(q_i+q_j)-\tfrac14 s_{ij},
\end{align}
with
\[ s_{ij} =\sum_\alpha\bigl(|(V_\alpha)_{ij}|^2+|(V_\alpha)_{ji}|^2\bigr)
\leq q_i+q_j.
\]
Consequently,
\[
\nu_{q,3}\ge\langle e(\psi_\theta)\rangle 
\geq\tfrac14(\|u_i-u_j\|^2+q_i+q_j)
\geq\tfrac14(\|u_i\|^2+q_i)
=\tfrac14\langle e_i,G e_i\rangle.
\]
Since the initial unit vector $e_i$ was arbitrary, $G\leq4\,\nu_{q,3} I_n$, proving
\eqref{eq:four}.

}

\section{Bound from inside for a single jump}\label{app:inside}

{

The excluded lens of Remark~\ref{rem:lens} was derived for a normal jump.
Here we show that the same exclusion holds for an arbitrary single jump
operator, and then construct an eigenvalue-free disk inside the lens.
Throughout this appendix,
\[
\D_A(X)=AXA^\dagger-\frac12(GX+XG),
\qquad G=A^\dagger A,
\qquad A\in M_n(\mathbb C),
\]
with no additional Hamiltonian term in this representation. No normality
assumption is imposed on $A$ or on the superoperator $\D_A$.
Write
\[
g_{\min}=\lambda_{\min}(G),\qquad
 g_{\max}=\lambda_{\max}(G),\qquad
 \Delta=g_{\max}-g_{\min},\qquad
 m=\frac{g_{\min}+g_{\max}}2.
\]
Since $G$ is Hermitian and positive semidefinite, its numerical range is
$W(G)=[g_{\min},g_{\max}]$~\cite{HornJohnson2013}.
The enclosures below depend only on these two endpoints.

\begin{proposition}%[Excluded lens for an arbitrary single jump]
\label{prop:appc-lens}
Define the open set
\begin{equation}\label{eq:appc-lens}
E=\{z\in\mathbb C:|z+g_{\min}|<g_{\min}\}
\cap
\{z\in\mathbb C:|z+m|<\sqrt{g_{\min}g_{\max}}\}.
\end{equation}
Then
\begin{equation}\label{eq:appc-sharp-enclosure}
\spec(\D_A)\subseteq D(-g_{\max},g_{\max})\setminus E.
\end{equation}
The lens $E$ is nonempty if and only if $g_{\max}<9g_{\min}$.
%Only the open lens is excluded; its boundary may contain eigenvalues.
\end{proposition}

\begin{proof}
Choose an orthonormal basis that diagonalizes $G$, not necessarily $A$:
\[
G=\operatorname{diag}(x_1,\ldots,x_n),
\qquad x_j\in[g_{\min},g_{\max}].
\]
Let $\D_A(X)=\lambda X$ with $X\ne0$, and write
$\lambda=-\Gamma+i\omega$. Since $\D_A$ generates a completely positive
trace-preserving semigroup, $\Gamma\ge0$.
The eigenvalue equation gives
\[
AXA^\dagger=\lambda X+\frac12(GX+XG).
\]
For the Hilbert--Schmidt norm $\|X\|_2^2=\tr(X^\dagger X)$, cyclicity of
the trace yields
\begin{align*}
\|AXA^\dagger\|_2^2
 &=\tr\bigl((AXA^\dagger)^\dagger AXA^\dagger\bigr)
 =\tr(X^\dagger GXG)
 =\sum_{j,k}x_jx_k|X_{jk}|^2,\\
\left\|\lambda X+\frac12(GX+XG)\right\|_2^2
 &=\sum_{j,k}\left|\lambda+\frac{x_j+x_k}{2}\right|^2|X_{jk}|^2.
\end{align*}
Consequently, with
\[
w_{jk}=\frac{|X_{jk}|^2}{\|X\|_2^2},
\qquad w_{jk}\ge0,\qquad \sum_{j,k}w_{jk}=1,
\]
we obtain the exact identity
\begin{equation}\label{eq:appc-weighted-identity}
|\lambda|^2
=\sum_{j,k}w_{jk}
\left[(x_j+x_k)\Gamma-\frac14(x_j-x_k)^2\right].
\end{equation}
This argument uses an eigenoperator for each eigenvalue; it does not
require diagonalizability of $\D_A$.

Each term in square brackets is at most $2g_{\max}\Gamma$, so
$|\lambda|^2\le2g_{\max}\Gamma$, equivalently
$|\lambda+g_{\max}|\le g_{\max}$. This proves the outer-disk inclusion.
For the lower bound, if $x,y\in[g_{\min},g_{\max}]$ and $\Gamma\ge0$, then
\begin{equation}\label{eq:appc-scalar-lower}
(x+y)\Gamma-\frac14(x-y)^2
\ge
\min\left\{2g_{\min}\Gamma,
 (g_{\min}+g_{\max})\Gamma-\frac14\Delta^2\right\}.
\end{equation}
Indeed, with $d=|x-y|\in[0,\Delta]$, one has $x+y\ge2g_{\min}+d$.
The function $(2g_{\min}+d)\Gamma-d^2/4$ is concave in $d$, so its
minimum on $[0,\Delta]$ is attained at an endpoint.
Averaging \eqref{eq:appc-scalar-lower} with the weights $w_{jk}$ gives
\begin{equation}\label{eq:appc-spectral-lower}
|\lambda|^2\ge
\min\left\{2g_{\min}\Gamma,
 (g_{\min}+g_{\max})\Gamma-\frac14\Delta^2\right\}.
\end{equation}
On the other hand, for $z=-\Gamma+i\omega$,
\begin{align*}
|z+g_{\min}|^2-g_{\min}^2
 &=|z|^2-2g_{\min}\Gamma,\\
|z+m|^2-g_{\min}g_{\max}
 &=|z|^2-(g_{\min}+g_{\max})\Gamma+\frac14\Delta^2.
\end{align*}
Thus $z\in E$ would make $|z|^2$ strictly smaller than both terms in
the minimum in \eqref{eq:appc-spectral-lower}. No eigenvalue can therefore
belong to $E$. In particular, $0\notin E$.

If $g_{\min}=0$, the first open disk in \eqref{eq:appc-lens} is empty.
If $g_{\min}>0$, the two open disks intersect exactly when the distance
between their centers is smaller than the sum of their radii:
\[
\frac{g_{\max}-g_{\min}}2
<g_{\min}+\sqrt{g_{\min}g_{\max}}.
\]
Writing $a=\sqrt{g_{\min}}$ and $b=\sqrt{g_{\max}}$, this is equivalent
to $(b-3a)(b+a)<0$, or $g_{\max}<9g_{\min}$.
\end{proof}

A nontrivial inner disk exists in this construction precisely when
$g_{\max}<9g_{\min}$. Under this assumption put
$a=\sqrt{g_{\min}}$, $b=\sqrt{g_{\max}}$ and
\begin{equation}\label{eq:cin}
C_{\rm in}=-\frac{5a^2-2ab+b^2}{4},
\qquad
R_{\rm in}=\frac{3a^2+2ab-b^2}{4}>0.
\end{equation}

\begin{proposition}%[Inner disk and ring]
\label{prop:ring}
Suppose $g_{\max}<9g_{\min}$. Then
\[
\operatorname{int}D(C_{\rm in},R_{\rm in})\subseteq E,
\]
so this open disk contains no eigenvalue of $\D_A$. In particular,
\begin{equation}\label{eq:ring1}
\spec(\D_A)\subseteq
\mathcal R:=D(-g_{\max},g_{\max})
\setminus\operatorname{int}D(C_{\rm in},R_{\rm in}),
\end{equation}
and $0\in\mathcal R$.
\end{proposition}

\begin{proof}
Set $p=-C_{\rm in}$. Since $b\ge a$, we have
\[
p-a^2=\frac{(b-a)^2}{4}\ge0,
\qquad
m-p=\frac{(b-a)(b+3a)}4\ge0.
\]
Moreover,
\[
R_{\rm in}+(p-a^2)=a^2,
\qquad
R_{\rm in}+(m-p)=ab.
\]
If $|z+p|<R_{\rm in}$, the triangle inequality therefore gives
\[
|z+a^2|<a^2,
\qquad |z+m|<ab,
\]
which is exactly $z\in E$. Proposition~\ref{prop:appc-lens} now yields \eqref{eq:ring1}. Finally,
\[
|C_{\rm in}|-R_{\rm in}=\frac12(a-b)^2\ge0,
\]
so the open inner disk does not contain the origin.
\end{proof}

}

\FloatBarrier
\bibliographystyle{unsrt}
\bibliography{spectral_references_13}
\end{document}